\documentclass[11pt]{article}

\PassOptionsToPackage{table}{xcolor}

\usepackage[preprint]{acl}

\usepackage{times}
\usepackage{latexsym}
\usepackage[T1]{fontenc}
\usepackage[utf8]{inputenc}
\usepackage{microtype}
\usepackage{inconsolata}
\usepackage{graphicx}

\usepackage{booktabs}       % Professional tables
\usepackage{amsmath}         % Math environments
\usepackage{amssymb}         % Math symbols
\usepackage{enumitem}        % List formatting (nosep, leftmargin)
\usepackage{xspace}          % Smart spacing after macros
\usepackage{multirow}        % Multi-row table cells
\usepackage{tabularx}        % Auto-balanced column widths (X column)
\usepackage{subcaption}      % Subfigures
\usepackage{colortbl}        % Cell/row coloring (xcolor[table] loaded above via \PassOptionsToPackage)
\usepackage{makecell}
\usepackage[most]{tcolorbox} 
\usepackage{mathtools}
\usepackage{float}
\DeclareFontShape{T1}{ptm}{m}{scit}{<-> ssub * ptm/m/sc}{}

\newcommand{\cara}{\textsc{cara}\xspace}
\newcommand{\carahyb}{\textsc{cara-hyb}\xspace}
\newcommand{\caranli}{\textsc{cara-nli}\xspace}
\newcommand{\carasim}{\textsc{cara-sim}\xspace}
\newcommand{\carallm}{\textsc{cara-llm}\xspace}
\newcommand{\gdp}{\textsc{gdp}\xspace}

\newcommand{\eg}{e.g.,\xspace}

\title{The Consistency Illusion: How Multi-Agent Debate Hides Reasoning Misalignment}

\author{
 \textbf{Xiaoyang Wang\textsuperscript{1}},
 \textbf{Christopher C. Yang\textsuperscript{1}}
\\
 \textsuperscript{1}Drexel University
\\
 \texttt{xw388@drexel.edu}, \texttt{chris.yang@drexel.edu}
}

\begin{document}
\maketitle

\begin{abstract}
Multi-agent LLM systems for medical question answering treat consensus as a reliability signal: if multiple agents agree on an answer, it is presumed trustworthy.
However, answer-level consensus does not entail reasoning-level alignment.
We introduce \textbf{CARA} (Cross-Agent Reasoning Alignment), a family of automated metrics that measure whether agents who agree on an answer also agree on the reasoning.
Applying \cara to a standard debate system on two medical QA benchmarks (MedQA-USMLE and MedThink-Bench), we identify the \textbf{consistency illusion}: a failure mode where debate reduces detectable contradictions between agents while simultaneously decreasing the semantic similarity of their reasoning chains; agents appear to agree more but reason less consistently.
To improve this misalignment, we propose the \textbf{Grounded Debate Protocol} (\gdp), a prompt-level intervention that requires agents to commit to named medical facts and take explicit stances on other agents' claims.
\gdp produces large, consistent alignment improvements (Cohen's $d = +1.43$ to $+1.99$) across two datasets and two backbone models, without adding LLM calls or modifying system architecture.
Our results motivate cross-agent reasoning alignment as a quantity to audit alongside accuracy in safety-critical domains.
\end{abstract}

% ============================================================
% §1 Introduction
% ============================================================
\section{Introduction}
\label{sec:intro}

Multi-agent LLM systems are increasingly used for medical question answering.
Frameworks such as MedAgents~\citep{tang2024medagents}, MDAgents~\citep{kim2024mdagents}, and MCC~\citep{sun2026model} deploy several LLM instances that debate, consult, and vote to answer clinical questions, and report accuracy gains over single-agent baselines on benchmarks such as MedQA.
Because the dominant evaluation pipeline primarily reports macroscopic answer accuracy under majority vote, the underlying \emph{trust assumption} is straightforward: if multiple agents independently converge on the same answer, that answer is reliable.

\begin{figure}[t]
\centering
\includegraphics[width=\columnwidth]{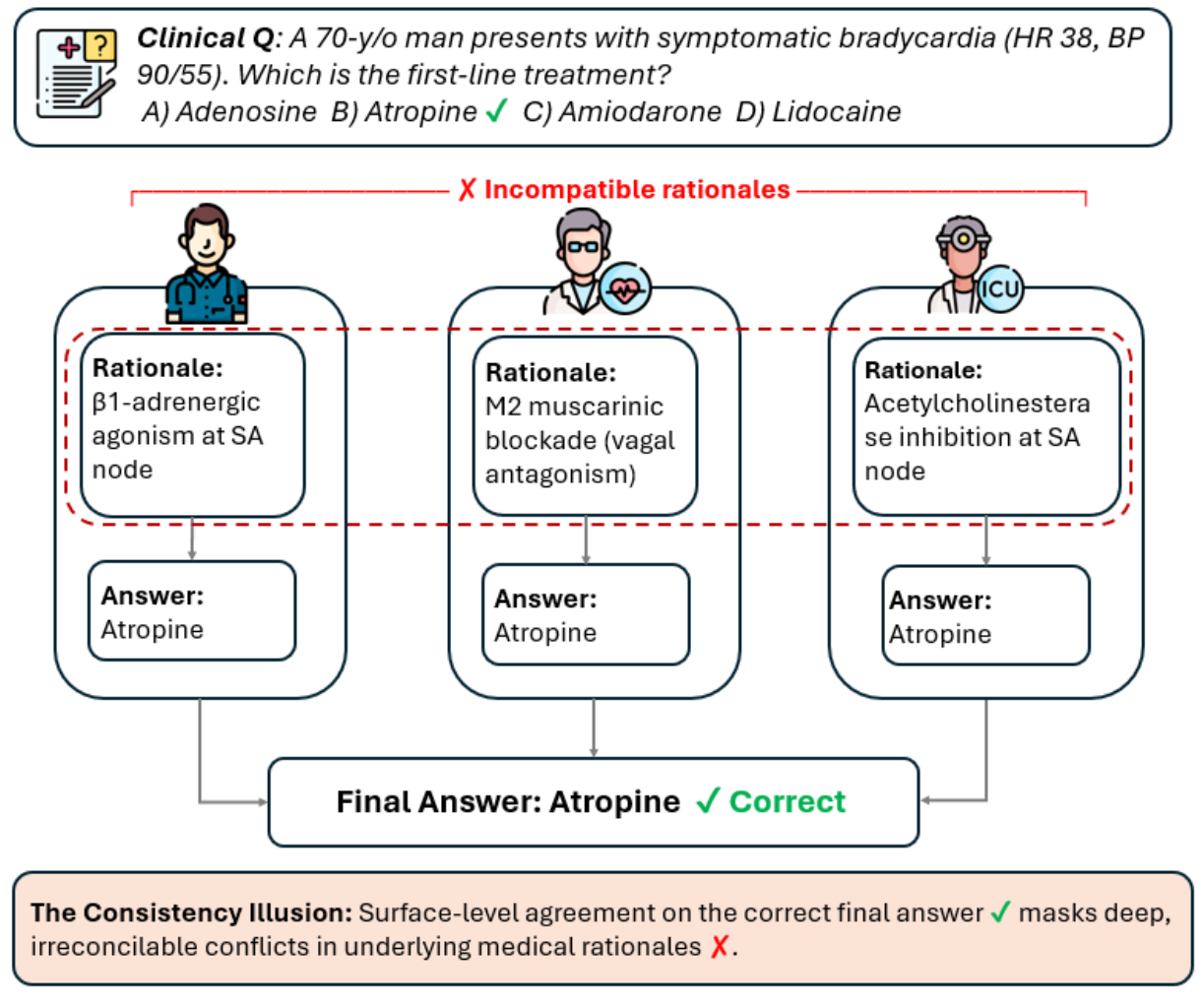}
\caption{The \textbf{consistency illusion}: three agents independently agree on the clinically correct answer (atropine for symptomatic bradycardia), yet their rationales invoke three mutually exclusive pharmacological targets.
Empirical 2D visualization on D2 (MedThink-Bench) appears in Appendix~\ref{app:illusion_2d}.}
\label{fig:illusion_concept}
\end{figure}

We argue this trust assumption is incomplete: answer-level consensus does not entail \emph{reasoning-level} alignment.
Figure~\ref{fig:illusion_concept} illustrates the gap on a high-stakes case: three agents independently agree that atropine is the first-line treatment for symptomatic bradycardia, yet their underlying rationales invoke three mutually exclusive pharmacological targets ($\beta_1$-adrenergic agonism, $M_2$-muscarinic blockade, and acetylcholinesterase inhibition).
The agents reach the correct answer through medically incompatible reasoning; we term this pattern the \textbf{consistency illusion}.
This paper therefore measures \textbf{cross-agent reasoning alignment}: whether agents that converge on an answer also align in the reasoning that produced it.
This question is orthogonal to both macroscopic answer accuracy and individual, single-agent reasoning faithfulness.

This phenomenon has gone undetected because existing evaluations measure something else.
Medical multi-agent systems report accuracy and answer agreement, while reasoning-faithfulness methods were built to audit a \emph{single} agent's trace rather than alignment \emph{across} agents; to our knowledge, no prior method measures whether agents that agree on an answer also agree on the reasoning behind it. Recent theoretical critiques of multi-agent deliberation~\citep{choi2025martingale,zhu2025medagentboard} characterize debate dynamics as a voting-driven martingale rather than genuine synthesis.
They leave open a complementary question central to safety-critical settings: when agents \emph{do} agree, does the reasoning behind that agreement hold up?

To answer this question, we introduce \textbf{\cara} (\textbf{C}ross-\textbf{A}gent \textbf{R}easoning \textbf{A}lignment), a family of automated metrics that decompose each agent's response into reasoning steps and score pairwise alignment among agents that share an answer, using a hybrid of NLI-based contradiction detection and embedding similarity.
\cara introduces no additional LLM calls and operates post hoc on traces produced by ordinary system inference.
Applying \cara to a standard three-agent debate system on MedQA-USMLE ($N{=}500$) and MedThink-Bench ($N{=}499$), we find that debate reduces explicit contradictions between agents (CR$\,\downarrow$) while \emph{simultaneously} reducing the semantic similarity of their reasoning chains (SIM$\,\downarrow$).
Debate produces surface harmony while reasoning diverges further: the empirical signature of the consistency illusion.
The effect is pronounced in open-ended reasoning, where the unconstrained reasoning space gives divergence room to surface: on MedThink-Bench it is large and replicates across both backbones (Cohen's $d{=}{-}0.30$ for Qwen, ${-}1.32$ for Llama-3), and it is muted on closed-form multiple-choice (MedQA), consistent with the narrower reasoning surface.

To test whether this failure is fundamental to multi-agent debate or merely a protocol-level deficiency, we propose the \textbf{Grounded Debate Protocol} (\gdp), a prompt-level intervention requiring each reasoning step to pair a \textsc{Claim} with a named \textsc{Ground} and forcing explicit \textsc{Stance} statements toward other agents' claims.
\gdp produces large, consistent alignment gains across all three conditions tested (Cohen's $d{=}{+}1.43$ on D1 Qwen, ${+}1.62$ on D2 Qwen, and ${+}1.99$ on D2 Llama-3).
Together, our results identify a previously unmeasured failure mode in multi-agent medical QA, name and quantify it, and demonstrate that lightweight prompt-level grounding substantially improves cross-agent reasoning alignment.
We release the code and computed \cara scores publicly\footnote{Code and data: \url{https://anonymous.4open.science/r/consistency-illusion-code-3629/}}.

% ============================================================
% §2 Related Work
% ============================================================
\section{Related Work}
\label{sec:related_work}

\paragraph{Multi-agent systems for medical QA.}
A growing family of medical MAS---MedAgents~\citep{tang2024medagents}, MDAgents~\citep{kim2024mdagents}, MCC~\citep{sun2026model}, TeamMedAgents~\citep{mishra2025teammedagents}, the foundational debate framework of~\citet{du2024debate}, and clinical-simulation systems like Agent Hospital~\citep{li2024agenthospital} and AgentClinic~\citep{fan2025agentclinic}---deploy multiple LLM instances that debate, consult, or vote on medical questions, reporting accuracy gains over single-agent baselines on benchmarks such as MedQA.
A common thread unites these architectures: they treat multi-agent answer-level consensus as a proxy for reliable clinical reasoning.
None, however, audits whether agents that converge on the same option actually share consistent underlying rationales.

\paragraph{Critiques of multi-agent deliberation.}
A growing body of work questions whether multi-agent debate improves reasoning beyond what voting or repeated sampling already provides.
\citet{choi2025martingale} prove formally that debate dynamics form a martingale---gains are attributable to majority voting, not deliberation---a conclusion reinforced empirically by \citet{zhu2025medagentboard}, \citet{li2024moreagents}, \citet{wang2024rethinking}, \citet{tran2026singleagent}, and \citet{zhang2025overvaluing}.
Behavioral studies document specific failure modes: agents may reach consensus by copying explanations~\citep{consens2025}, switch answers under sycophantic pressure~\citep{yao2026sycophancy}, or degrade in multi-turn dialogue~\citep{laban2026multiturn}.
Most closely related, the problem drift literature~\citep{becker2026drift} measures how debate performance degrades over turns; \cara{} asks a complementary question: when agreement does emerge, is the underlying reasoning aligned?
Prior critiques attack the debate \emph{process}; our work attacks an assumption about the debate \emph{outcome}---that surface consensus implies reasoning alignment.

\paragraph{Evaluation of multi-agent reasoning.}
A distinct evaluation lineage measures whether reported rationales support or reflect a single agent's answer.
\citet{lanham2023faithfulness} introduced CoT truncation and mistake-injection probes; \citet{matton2025walkthetalk} extended these to medical QA, and \citet{afolabi2025cot} show that reasoning steps often do not causally drive predictions.
Automated metrics include ROSCOE~\citep{golovneva2023roscoe}, NLI-based faithfulness checking~\citep{falke2019ranking}, and CC-SHAP~\citep{parcalabescu2024ccshap}, but these are designed for single-trace verification of individual model outputs.
Even the methods that do apply to multi-agent settings verify each trace in isolation; to our knowledge, none measures cross-agent reasoning alignment among agents that share an answer.
How multi-agent aggregation affects collective reasoning coherence has therefore remained uncharacterized.

% ============================================================
% §3 CARA Metric
% ============================================================
\section{CARA: Measuring Reasoning Alignment}
\label{sec:cara}

\cara{} measures whether agents that agree on an answer also agree on the reasoning behind it---a quantity that answer accuracy and single-agent faithfulness leave unmeasured.
It is \emph{orthogonal} to correctness: rather than checking whether a rationale is medically right, it asks whether agents who reach the same answer reach it for compatible reasons.

\subsection{Agreement Set}
\label{sec:cara:setup}

We run $N$ agents on a clinical question $q$.
Each agent $A_i$ produces an answer $a_i$ and a reasoning chain $R_i = (r_i^1, \dots, r_i^{K_i})$ of $K_i$ steps.
The system answer $\hat{a}$ is the majority vote over $\{a_i\}$.
We define the \textbf{agreement set} as the agents who voted for $\hat{a}$:
\begin{equation}
  \mathcal{S}(q) = \{i : a_i = \hat{a}\}.
  \label{eq:agreement_set}
\end{equation}
We compute \cara{} only when $|\mathcal{S}(q)| \geq 2$.
This restriction is the point: it isolates the safety-critical case where surface agreement may hide divergent reasoning.

\subsection{Step-Level Alignment}
\label{sec:cara:scoring}

We split each rationale $R_i$ into steps using a deterministic text-segmentation pipeline.
For each ordered pair $(A_i, A_j)$ in the agreement set with $i \neq j$, we build an alignment matrix $\mathbf{M}_{ij} \in \mathbb{R}^{K_i \times K_j}$ with entries $\mathbf{M}_{ij}[k,l] = \mathrm{align}(r_i^k, r_j^l)$.

% We use four variants of the alignment function (Table~\ref{tab:cara-variants}, Appendix~\ref{app:variants}).
Our metric is \carahyb{}; \caranli{} and \carasim{} are its two components, reported separately in Table~\ref{tab:cara-variants} to expose what drives alignment.
\textbf{\caranli} returns an NLI label in $\{-1, 0, +1\}$ for contradiction, neutral, or entailment.
\textbf{\carasim} returns the cosine similarity of step embeddings.
\textbf{\carahyb} combines both, letting contradictions override similarity:
\begin{equation}
  \mathrm{align}_{\mathrm{hyb}}(r_i^k, r_j^l) =
  \begin{cases}
    -1 & \text{if } P_{\mathrm{NLI}} > \tau, \\
    \cos(\mathbf{e}_i^k, \mathbf{e}_j^l) & \text{otherwise,}
  \end{cases}
  \label{eq:align_hyb}
\end{equation}
where $P_{\mathrm{NLI}}$ is the NLI contradiction probability for $(r_i^k, r_j^l)$, $\mathbf{e}_i^k$ is the L2-normalized embedding of $r_i^k$, and $\tau$ is the contradiction threshold.
This hybrid addresses the known insensitivity of embeddings to negation~\citep{marelli-etal-2014-sick}.
% A fourth variant, \textbf{\carallm}, uses GPT-4o as an LLM-judge for calibration (Section~\ref{sec:exp:carallm}).
For validation, \textbf{\carallm{}} uses GPT-4o as an LLM judge to calibrate \carahyb{} against holistic human-style judgments, reported in Section~\ref{sec:exp:carallm}.

\subsection{Question-Level Aggregation and Diagnostics}
\label{sec:cara:aggregation}

\paragraph{Best-match alignment.}
Because agents produce different numbers of steps ($K_i \neq K_j$), alignment is asymmetric.
We define \textbf{best-match} alignment by matching each step in $A_i$ to its best match in $A_j$:
\begin{equation}
  \mathrm{match}_j(r_i^k) = \max_{l \in \{1,\dots,K_j\}} \mathrm{align}(r_i^k,\, r_j^l).
  \label{eq:bestmatch}
\end{equation}

\paragraph{Symmetrized pairwise score.}
The pairwise \cara{} score symmetrizes by averaging both directions:
\begin{multline}
  \mathrm{CARA}(i, j) = \tfrac{1}{2}\Big[
    \tfrac{1}{K_i}\!{\textstyle\sum_{k=1}^{K_i}}\, \mathrm{match}_j(r_i^k) \\
    + \tfrac{1}{K_j}\!{\textstyle\sum_{l=1}^{K_j}}\, \mathrm{match}_i(r_j^l)
  \Big].
  \label{eq:pairwise}
\end{multline}

\paragraph{Question-level aggregation.}
The question-level score averages over all pairs in the agreement set:
\begin{multline}
  \mathrm{CARA}(q) = \frac{2}{|\mathcal{S}(q)|\,(|\mathcal{S}(q)|{-}1)} \\
    \times \sum_{\substack{i < j \\ i,j \in \mathcal{S}(q)}} \mathrm{CARA}(i, j).
  \label{eq:question}
\end{multline}
Variants with negative range (\caranli, \carahyb) are rescaled to $[0,1]$ via $(\mathrm{CARA}{+}1)/2$.
Corpus-level scores are the mean of $\mathrm{CARA}(q)$ over questions with $|\mathcal{S}(q)| \geq 2$.

\paragraph{Contradiction Rate (CR).}
We also report the \textbf{Contradiction Rate}: the fraction of best-matched step pairs in the agreement set flagged as contradictions by the NLI filter.
\begin{equation}
  \mathrm{CR}(q) = \frac{\left|\left\{(i, j, k) : i \neq j, \ \mathrm{match}_j(r_i^k) = -1\right\}\right|}{\sum_{i \in \mathcal{S}(q)} (|\mathcal{S}(q)| - 1) K_i}.
  \label{eq:cr}
\end{equation}
True alignment shows up as semantic similarity rising (\carasim$\uparrow$) with contradictions falling (CR$\downarrow$).
When the two diverge---contradictions fall but similarity also falls---we see the consistency illusion.
% ============================================================
% §4 GDP Protocol
% ============================================================
\section{Grounded Debate Protocol: A Protocol-Level Intervention}
\label{sec:gdp}

\subsection{Motivation and Protocol Design}
\label{sec:gdp-design}

\cara{} shows that free-form debate can produce answer agreement without aligned reasoning, a result we establish in Section~\ref{sec:results}.
\gdp{} tests whether this failure is an unresolvable feature of language-model interaction or merely a protocol-level deficiency.
Standard debate prompts allow three failure modes: vague clinical claims; sycophantic answer adoption without engaging with peer evidence~\citep{yao2026sycophancy}; and explanation copying with topic drift~\citep{consens2025,becker2026drift}.
The common root is that nothing in standard prompts requires agents to commit to specific reasoning steps or substantively engage with other agents' claims.

To address this, \gdp{} enforces a structured, prompt-level communication protocol.
It requires each agent to structure its output into up to three machine-parseable fields:
\begin{itemize}[leftmargin=*,nosep]
\item \textsc{Claim}: A singular, atomic, and falsifiable clinical assertion.
\item \textsc{Ground}: a named medical fact, mechanism, or guideline that supports the claim (\eg{} disease, drug target, receptor pathway).
\item \textsc{Stance} (\emph{debate rounds only}): one of \{\textsc{Agree}, \textsc{Disagree}, \textsc{Extend}\} toward a specific claim from another agent, with a one-sentence justification; a \textsc{Disagree} requires a counter-\textsc{Ground}.
\end{itemize}
Agents emit \textsc{Claim}+\textsc{Ground} pairs in the independent round ($r_0$) and add \textsc{Stance} in the debate round ($r_1$). 
% In $r_1$, each agent sees the other two agents' $r_0$ \textsc{Claim}+\textsc{Ground} (never its own), revises its answer, and adds \textsc{Stance} statements toward their claims.
An anti-sycophancy rule instructs each agent to switch answers between $r_0$ and $r_1$ only after seeing a more compelling \textsc{Ground} from another agent or finding a factual error in its own reasoning.
A worked clinical example contrasting an unstructured debate log with a synchronized \gdp{} trace is provided in Appendix~\ref{app:gdp_example}.

\subsection{Integration Properties}
\label{sec:gdp-integration}

\gdp{} is a prompt-level change: the \textsc{Claim}/\textsc{Ground}/\textsc{Stance} format is added through the system message, with agent routing and the number of LLM calls unchanged.
The only overhead is output length: tokens grow by ${\sim}9\%$ from the explicit field markers.
Applied to the symmetric debate of \citet{du2024debate}, this gives M3-GDP, which we evaluate in Section~\ref{sec:exp:design}.
Traces are parsed post-hoc by regex on the field markers; this parsing supports logging and analysis only, as \cara{} operates on the raw response text regardless of parse outcome.

% ============================================================
% §5 Experimental Setup
% ============================================================
\section{Experimental Setup}
\label{sec:experiments}

\subsection{Study Design}
\label{sec:exp:design}

We evaluate whether debate-induced answer consensus is accompanied by reasoning alignment among agents that converge on the same answer.
Table~\ref{tab:systems} summarizes our experimental matrix of three systems, each with three agents under majority voting:
\textbf{M3} is symmetric one-round debate~\citep{du2024debate}, producing pre-debate ($r_0$) and post-debate ($r_1$) traces;
\textbf{M4} is an independent-vote control with no cross-agent communication, specifically isolating the sampling-and-voting contribution from debate~\citep{choi2025martingale};
\textbf{M3-GDP} keeps the M3 architecture and the same LLM-call budget but rewrites all prompts to require the \gdp \textsc{Claim}+\textsc{Ground}+\textsc{Stance} format (Section~\ref{sec:gdp}).
Full configurations and prompts are in Appendix~\ref{app:systems}.

\subsection{Datasets and Backbones}
\label{sec:exp:data}

\textbf{D1: MedQA-USMLE}~\citep{jin2020medqa} provides closed-form four-option USMLE-style multiple-choice questions; we sample $N{=}500$ stratified by exam stage with a fixed seed.
\textbf{D2: MedThink-Bench}~\citep{zhou2025automating} spans ten medical domains with $3$--$10$ answer options and expert-annotated reasoning trajectories (``Scoring Points''); we use $N{=}499$ of 500 available items (one excluded for incompatible option format).
D2's wider option space gives reasoning more room to diverge than D1.
Primary trials use Qwen~2.5 72B Instruct~\citep{qwen2025qwen25}; we replicate on D2 ($N{=}100$) with Llama~3.3 70B Instruct~\citep{grattafiori2024llama3} to test the effect-direction consistency rather than cross-backbone statistical significance.
% Serving environment, decoding, and GPU details are in Appendix~\ref{app:systems}.

\subsection{CARA Computation and Validation}
\label{sec:exp:cara}

\paragraph{\carahyb implementation.}
\carahyb decomposes each agent response into reasoning steps and computes pairwise alignment via two complementary signals: an NLI hard-filter on contradictions (DeBERTa-v3~\citep{laurer-etal-2024-less}) and a sentence-embedding cosine on non-contradictory step pairs (Stella~\citep{zhang2024stella}).
The pipeline runs post hoc with no additional LLM calls; model versions, threshold, decomposition rules, and runtime are in Appendix~\ref{app:carahyb_impl}.

\noindent\textbf{Agent--expert validation (\cara-GT).}\label{sec:exp:caragt}~For D2 only, \cara-GT applies the same pipeline with one agent replaced by the expert Scoring-Points reference, measuring agent-to-expert alignment and \emph{expert coverage}, detailed in Appendix~\ref{app:caragt}.

\noindent\textbf{LLM-judge calibration (\carallm).}\label{sec:exp:carallm}~\carallm validates \carahyb against GPT-4o-as-judge~\citep{openai2024gpt4o} on a sample of agent pairs: the scores show strict tercile monotonicity (and a moderate linear correlation), capturing complementary aspects of alignment---step-level semantic overlap (\carahyb) vs.\ holistic logical coherence (GPT-4o judge).
Sample design, correlation values, and tercile statistics are in Appendix~\ref{app:carallm}.

\noindent\textbf{Human-expert validation.}\label{sec:exp:human}~Because reliable annotation of medical reasoning requires medically-trained experts, we validate on a deliberately focused set of $50$ agreement-set pairs rated by three independent annotators, with the full setup in Appendix~\ref{app:human_annotation}.
Inter-annotator agreement is high (Spearman $\rho \in [0.93, 0.99]$; $100\%$ within-1 agreement), and aggregate ratings correlate positively with \carahyb (Spearman $\rho{=}0.26$) and \carasim ($\rho{=}0.31$), with tercile means rising monotonically ($4.00/4.65/4.71$ for low/mid/high \carahyb strata).
The bounded magnitude reflects range restriction within the agreement set ($\carahyb \in [0.76, 0.98]$), paralleling the \carallm pattern.

\subsection{Statistical Analysis}
\label{sec:exp:stats}

We report Cohen's $d$ with $95\%$ percentile bootstrap CIs ($B{=}10{,}000$, seed~$42$); $|d|{>}0.8$ is Tier A, $0.5{<}|d|{\leq}0.8$ Tier B, and $0.2{<}|d|{\leq}0.5$ Tier C.
Accuracy comparisons use continuity-corrected McNemar tests.
Questions with fewer than two agents choosing the same answer are undefined; we report this undefined rate per system and check survivorship bias by imputing the undefined cases. Per-cell counts are in Appendix~\ref{app:undefined}, and the full imputation in Appendix~\ref{app:sensitivity} and Section~\ref{sec:results:survivorship}.

% ============================================================
% §6 Results
% ============================================================
\section{Results}
\label{sec:results}

We report \carahyb, \carasim, and CR for all three systems on D1 and D2 with Qwen 2.5 72B, plus a cross-backbone replication on D2 ($N{=}100$) with Llama 3.3 70B.
All comparisons are within-dataset and within-backbone; we do not compare absolute \cara{} values across datasets.
% Bootstrap confidence intervals use the percentile method with $B{=}10{,}000$ iterations and seed $42$.
% Results come in three layers: main alignment with survivorship robustness (\S\ref{sec:results:main}, \S\ref{sec:results:gdp}), cross-backbone validation (\S\ref{sec:results:backbone}), and supplementary external validation (\S\ref{sec:results:caragt}, \S\ref{sec:results:accuracy}).
\begin{table*}[ht]
\centering
\small
\setlength{\tabcolsep}{8pt} 
\definecolor{rowBest}{HTML}{81C784}   % strong green: M3-GDP r1 (highest alignment)
\definecolor{rowGood}{HTML}{C8E6C9}   % pale green:   M3-GDP r0 (format effect only)
\definecolor{rowMid}{HTML}{F5F5F5}    % light gray:   M4 r0, M3 r0 (independent baselines)
\definecolor{rowWorst}{HTML}{EF9A9A}  % medium red:   M3 r1 (consistency illusion)
% Legacy aliases (kept for backward compat with caption references)
\colorlet{gdpgreen}{rowBest}
\colorlet{m3red}{rowWorst}
\begin{tabular}{@{}llcccccc@{}}
\toprule
& & \multicolumn{3}{c}{\textbf{D1 (MedQA, $N{=}500$)}} & \multicolumn{3}{c}{\textbf{D2 (MedThink, $N{=}499$)}} \\
\cmidrule(lr){3-5} \cmidrule(l){6-8}
\textbf{System} & \textbf{Round} & \textbf{HYB} & \textbf{SIM} & \textbf{CR} & \textbf{HYB} & \textbf{SIM} & \textbf{CR} \\
\midrule
\rowcolor{rowBest}  M3-GDP & r1 (debate)  & \textbf{0.952} & \textbf{0.912} & 0.098 & \textbf{0.953} & \textbf{0.914} & 0.127 \\
\rowcolor{rowGood}  M3-GDP & r0 (indep.)  & 0.912 & 0.835 & 0.117 & 0.912 & 0.836 & 0.137 \\
\rowcolor{rowMid}   M4     & r0 (indep.)  & 0.895 & 0.801 & 0.108 & 0.895 & 0.801 & 0.161 \\
\rowcolor{rowMid}   M3     & r0 (indep.)  & 0.895 & 0.801 & 0.104 & 0.896 & 0.801 & 0.144 \\
\rowcolor{rowWorst} M3     & r1 (debate)  & 0.892 & 0.787 & 0.035$^{\dagger}$ & 0.881 & 0.769 & 0.054$^{\dagger}$ \\
\bottomrule
\end{tabular}
\caption{\carahyb, \carasim, and CR per cell on D1 and D2.
Higher HYB and SIM, and lower CR, indicate better alignment.
\colorbox{gdpgreen}{Green row} = M3-GDP r1 (highest alignment); \colorbox{m3red}{Red row} = M3 r1 (lowest, consistency illusion).
Per-cell HYB SDs are $0.021$--$0.059$.
$N$ ranges $424$--$500$ (D1) and $428$--$499$ (D2) due to undefined questions.
$^{\dagger}$M3~r1 CR is deflated by survivorship bias (Section~\ref{sec:results:survivorship}).}
\label{tab:main_results}
\end{table*}
\subsection{Main Results and Dual-Layer Decomposition}
\label{sec:results:main}

The rank ordering is identical on D1 and D2: M3-GDP~r1 is the highest cell ($0.952$/$0.953$), and M3~r1 is the \emph{lowest} ($0.892$/$0.881$)---standard debate alignment falls below independent voting.

For M3 from r0 to r1, \emph{both} CR and SIM decrease (CR $0.104{\to}0.035$ on D1, $0.144{\to}0.054$ on D2; SIM $0.801{\to}0.787$ on D1, $0.801{\to}0.769$ on D2); fewer contradictions with less semantic alignment is the \emph{consistency illusion} (mechanism in Section~\ref{sec:analysis}).

In contrast, \gdp{} debate produces the genuine-alignment signature.From r0 to r1, SIM \emph{rises} ($0.835{\to}0.912$ on D1; $0.836{\to}0.914$ on D2) while CR \emph{falls} ($0.117{\to}0.098$ on D1; $0.137{\to}0.127$ on D2). Both protocols reduce contradictions, but only \gdp{} simultaneously increases reasoning similarity---CR$\downarrow$ with SIM$\uparrow$, the true-alignment pattern defined in Section~\ref{sec:cara} and the mirror image of the illusion.
A D1 bootstrap test underscores that M3's low CR is \emph{anomalous}: \gdp~r1 CR exceeds M3~r1 ($\Delta{=}{+}0.063$, CI $[{+}0.042,{+}0.087]$), and M3~r1 CR falls even below independent voting ($\Delta{=}{-}0.072$ vs.\ M4, CI $[{-}0.089,{-}0.056]$)---standard debate suppresses visible contradictions without aligning reasoning.

\subsection{GDP Effect and the Consistency Illusion}
\label{sec:results:gdp}

\begin{figure}[t]
\centering
\includegraphics[width=0.76\columnwidth]{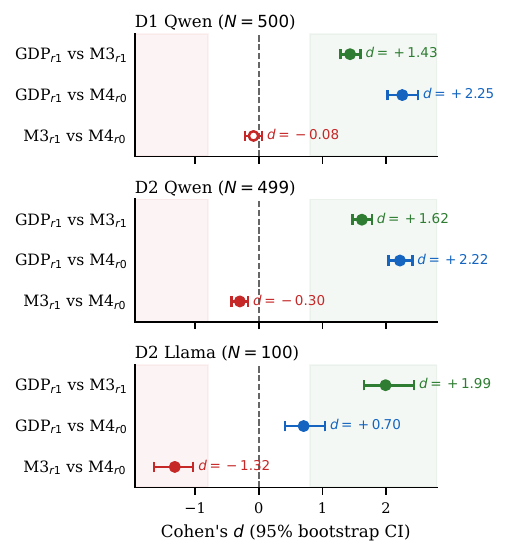}
\caption{Cohen's $d$ for three key pairwise comparisons, one panel per dataset--backbone.
Error bars: $95\%$ bootstrap CIs ($B{=}10{,}000$); filled markers exclude zero; shading marks Tier~A ($|d|{>}0.8$).
\gdp comparisons (top two rows) reach Tier~A across all conditions; the consistency-illusion comparison (bottom) is significantly negative on D2 (both backbones).}
\label{fig:gdp_forest}
\end{figure}

Figure~\ref{fig:gdp_forest} shows the three key pairwise comparisons as a forest plot of Cohen's $d$ across all three dataset--backbone combinations.

\gdp delivers a large alignment gain on both datasets: $d{=}{+}1.43$ on D1 and $d{=}{+}1.62$ on D2 against M3$_{r_1}$, rising to $d{=}{+}2.26$ and $d{=}{+}2.22$ against independent voting (all four CIs above zero).
Standard debate, in contrast, is alignment-neutral on D1 ($d{=}{-}0.08$, CI contains zero) and alignment-degrading on D2 ($d{=}{-}0.30$, CI $[{-}0.43,{-}0.17]$).
The D1--D2 gap tracks both question complexity and undefined-question selection bias.
The \gdp effect splits in two: \textsc{Claim}+\textsc{Ground} format alone (GDP$_{r_0}$ vs.\ M4$_{r_0}$) yields $d{=}{+}0.62$ on D1; adding \textsc{Stance}-mediated debate (GDP$_{r_1}$ vs.\ GDP$_{r_0}$) yields $d{=}{+}1.51$, $2.4{\times}$ the format component but only meaningful when there are structured claims to reference.
On survivorship-bias sensitivity,\label{sec:results:survivorship}
M3~r1 excludes $15.2\%$ (D1) and $14.2\%$ (D2) of questions where no majority emerged.
These exclusions are non-random---harder questions survive less often---so the sample skews toward easier cases.
Under worst-case imputation (\carahyb${=}0.50$, CR${=}1.0$ for every undefined question, both far beyond any observed value), M3~r1's adjusted \carahyb stays \emph{below} M4~r0 ($0.832$ vs.\ $0.895$ on D1; $0.827$ vs.\ $0.895$ on D2): the consistency-illusion conclusion is robust.
The CR finding, however, reverses ($0.182{>}0.108$ on D1; $0.188{>}0.161$ on D2), so we interpret CR alongside the undefined rate.

\subsection{Cross-Backbone Validation}
\label{sec:results:backbone}

We replicate all three systems on D2 ($N{=}100$) with Llama 3.3 70B Instruct.
All five key comparisons agree in sign across both backbones (Table~\ref{tab:backbone}): \gdp is Tier~A positive (Llama $d{=}{+}1.99$, Qwen $d{=}{+}1.62$), the illusion is negative (Llama $d{=}{-}1.32$, Tier~A; Qwen $d{=}{-}0.30$, Tier~C), standard debate degrades alignment r0$\to$r1 while \gdp debate improves it, and the format effect is positive.
The illusion is stronger on Llama partly because Llama's M3~r1 undefined rate ($2\%$) is far lower than Qwen's ($14\%$), so more hard cases survive into the measured sample.

\subsection{Agent--Expert Alignment (\cara-GT)}
\label{sec:results:caragt}

\begin{figure}[t]
\centering
\includegraphics[width=0.76\columnwidth]{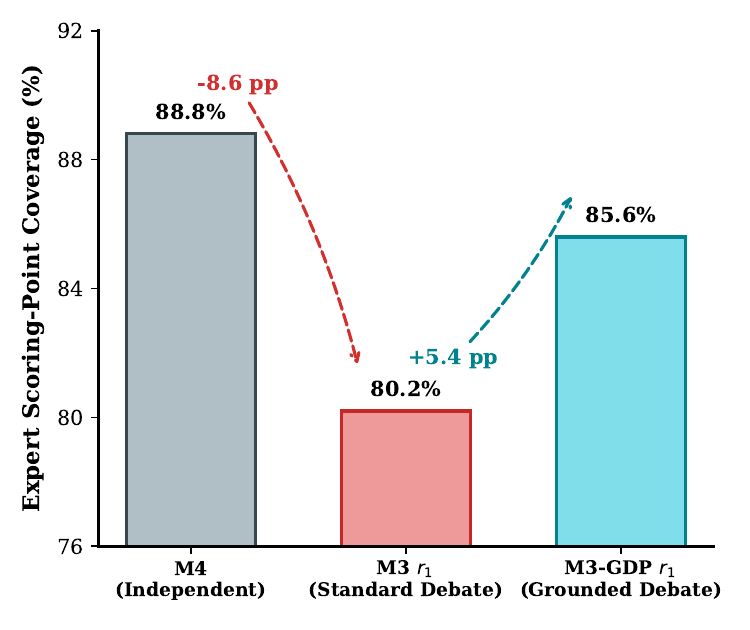}
\caption{Expert scoring-point coverage on D2: standard debate (M3~$r_1$) drops $8.6$pp from the independent-voting baseline; \gdp (M3-GDP~$r_1$) recovers $5.4$pp.}
\label{fig:caragt_coverage}
\end{figure}

\cara-GT replaces one agent with the expert Scoring-Points reference (D2 only; Section~\ref{sec:exp:caragt}).
It mirrors the inter-agent finding: \gdp moves agents \emph{toward} expert reasoning ($d{=}{+}0.32$, CI $[{+}0.20,{+}0.45]$), and standard debate moves them \emph{away} ($d{=}{-}0.21$, CI $[{-}0.34,{-}0.08]$).
Expert-coverage tells a sharper story (Figure~\ref{fig:caragt_coverage}): standard debate loses $8.6$pp of expert scoring-point coverage; \gdp recovers $5.4$pp.

\subsection{Accuracy}
\label{sec:results:accuracy}

\cara measures reasoning alignment, not answer correctness; majority-vote accuracy is reported separately.
Standard debate does not change accuracy on either dataset (D1 $\Delta{=}{+}0.4$pp, D2 $\Delta{=}0.0$pp; both $p{>}0.85$), consistent with the martingale characterization~\citep{choi2025martingale}.
\gdp shifts accuracy by ${-}2.4$pp on Qwen (both D1 and D2) and ${-}6.0$pp on Llama D2; none reaches significance at $\alpha{=}0.05$.
D2's lower absolute accuracy ($\sim 41\%$ baseline) reflects its $\sim 7.5$ average answer options versus D1's $4$, giving an ${\approx}3{\times}$ random-performance ratio.

\begin{table*}[!t]
\centering
\small
\begin{tabularx}{\textwidth}{ll X cccc}
\toprule
 & & & \multicolumn{2}{c}{\textbf{M3 r1}} & \multicolumn{2}{c}{\textbf{GDP r1}} \\
\cmidrule(lr){4-5} \cmidrule(lr){6-7}
\textbf{Severity} & \textbf{Mode} & \textbf{Definition} & \textbf{D1} & \textbf{D2} & \textbf{D1} & \textbf{D2} \\
\midrule
\textsc{Severe}   & FM1 & \textbf{Complementary reasoning}: different, non-contradictory reasoning paths. & 23 & 18 & \textbf{0} & \textbf{0} \\
                  & FM4 & \textbf{Sycophantic convergence}: agent adopts the majority answer with zero reasoning steps. & 11 & 15 & \textbf{0} & \textbf{0} \\
\addlinespace
\textsc{Moderate} & FM2 & \textbf{Granularity mismatch}: step-count ratio ${>}2$ (depth difference, not factual clash). & 12 & 15 & 5 & 2 \\
                  & FM3 & \textbf{Contradictory premises}: agents cite conflicting facts but converge on the correct answer. & 2 & 1 & 12 & 15 \\
                  & FM5 & \textbf{Terminology divergence}: different medical terms used for the same concept. & 2 & 1 & 0 & 0 \\
\addlinespace
\textsc{Mild}     & FM6 & \textbf{Partial overlap}: residual baseline misalignment without extreme signals. & 0 & 0 & 31 & 32 \\
\midrule
\multicolumn{3}{r}{\emph{Severe modes eliminated by \gdp{} (FM1 + FM4):}} & \textbf{34} & \textbf{33} & \textbf{0} & \textbf{0} \\
\bottomrule
\end{tabularx}
\caption{Failure mode taxonomy on the 50 lowest-\carahyb correct-answer cases per system, with definitions and counts on both datasets.
\gdp{} \emph{eliminates} both severe modes (FM1 + FM4: $34/50$ and $33/50$ under M3 $\to$ $0/50$ under \gdp{} on D1 and D2 respectively), shifting residual misalignment toward the mild baseline FM6. FM3's rise under \gdp{} is consistent with the CR analysis in §\ref{sec:results:main}: the structured format surfaces contradictions that vague free text would hide.}
\label{tab:fm_combined}
\end{table*}

% ============================================================
% §7 Analysis
% ============================================================
\section{Analysis}
\label{sec:analysis}

We answer four research questions about the consistency illusion, the mechanism by which \gdp improves it, and the generality of the findings.

\paragraph{RQ1: What mechanism produces the consistency illusion, and how does GDP improve it?}
\textbf{Answer}: \emph{Standard debate creates the illusion via two failures---contradiction smoothing without reasoning convergence, and sycophantic convergence---and \gdp reverses both via a format effect ($d{=}{+}0.62$) amplified by a \textsc{Stance}-mediated debate interaction effect ($d{=}{+}1.51$, $2.4{\times}$ the format component; extended derivation in Appendix~\ref{app:mechanism_extended}).}

Standard debate drives CR$\,\downarrow$+SIM$\,\downarrow$ (Section~\ref{sec:results:main}) via two mechanisms.
First, revision pressure removes contradictory steps without replacing them with shared reasoning---\emph{subtraction without alignment}---so CR drops while SIM also drops.
Second, agents adopt each other's answers without adopting the underlying reasoning, sometimes severely: M3~r1 fails consensus on $14$--$15\%$ of questions on both datasets.
\gdp reverses both via two complementary effects.
The \emph{format effect} alone, before any debate (GDP$_{\text{r0}}$ vs.\ M4$_{\text{r0}}$ on D1: SIM~$+0.034$, $d{=}{+}0.62$), shows that much of the apparent misalignment originates from underspecified output format rather than genuine reasoning disagreement.
The dominant \emph{debate interaction effect} ($d{=}{+}1.51$ on D1) operates because \textsc{Stance} statements force agents to share reasoning vocabulary and logical structure with their interlocutors, driving SIM up by $+0.077$ on D1 and $+0.078$ on D2 from r0 to r1.
Format is a precondition: without structured claims, \textsc{Stance} has no targets.

\paragraph{RQ2: Beyond CARA, what behavioral indicators support the illusion finding?}
\textbf{Answer}: \emph{Standard debate destabilizes consensus ${\sim}38{\times}$ more than independent voting on D1 and produces $5\%$ ``reasoning collapse'' zero-step agents on D2; \gdp eliminates both.}

Two independent behavioral indicators support the illusion finding without depending on agreement-set selection.
First, the \emph{undefined rate} (no majority emerges) jumps from $0.4\%$ before debate (M3~r0) to $15.2\%$ after (M3~r1) on D1 ($38{\times}$) and $5.6\%{\to}14.2\%$ on D2: debate itself manufactures disagreement; grounding removes it (M3-GDP~r1: $0.4\%$).
Second, \emph{reasoning collapse}---agents emitting zero extractable steps after debate, the behavioral signature of sycophantic convergence---affects $5.0\%$ of M3~r1 agreement-set memberships on D2 ($N{=}100$) vs.\ $0.3\%$ for M3-GDP~r1. \gdp's 
\textsc{Claim}+\textsc{Ground} requirement prevents both failure modes.

\paragraph{RQ3: Which failure modes does GDP structurally eliminate?}
\textbf{Answer}: \emph{\gdp eliminates FM1 (complementary reasoning) and FM4 (sycophantic convergence) on \emph{both} datasets, shifting residual misalignment toward the low-severity baseline FM6 (partial overlap, $62$--$64\%$ of remaining low-\carahyb cases).}

We examine the 50 lowest-\carahyb cases per system where each still reaches the correct answer, classifying by deterministic rules over CARA sub-metrics; the full classification rules are in Appendix~\ref{app:failure_modes}.
Table~\ref{tab:fm_combined} groups the six modes into \textsc{Severe} (FM1, FM4), \textsc{Moderate} (FM2, FM3, FM5), and \textsc{Mild} (FM6) tiers.

Four findings replicate across the two datasets.
First, \gdp \textbf{eliminates sycophantic convergence (FM4)}: $11{\to}0$ on D1, $15{\to}0$ on D2; the \textsc{Claim}+\textsc{Ground} requirement structurally prevents empty answer adoption.
Second, \gdp \textbf{eliminates complementary reasoning (FM1)}: $23{\to}0$ on D1, $18{\to}0$ on D2; structured grounding forces convergence on shared medical facts.
Third, residual misalignment under \gdp \textbf{shifts to FM6 (partial overlap)}: $31$/$50$ on D1, $32$/$50$ on D2 ($62$--$64\%$)---a low-severity baseline remaining after severe modes are removed.
Fourth, \textbf{FM3 contradictory premises rises under \gdp} ($2{\to}12$ on D1, $1{\to}15$ on D2), consistent with the CR analysis in Section~\ref{sec:results:main}: the structured format surfaces contradictions that vague free text would hide.

A qualitative case study (MQA\_0447 hypovolemic shock, where agents misstate SVR direction before self-correcting) is in Appendix~\ref{app:failure_modes}.
% Cross-dataset replication confirms that \gdp eliminates specific severe failure modes (FM1, FM4) across both datasets, while shifting residual cases toward the milder FM6 baseline.

\paragraph{RQ4: Do the findings generalize across datasets and backbones?}
\textbf{Answer}: \emph{Yes---the \gdp effect replicates with Tier-A magnitudes across all three conditions (D1 Qwen $d{=}{+}1.43$; D2 Qwen $d{=}{+}1.62$; D2 Llama $d{=}{+}1.99$), with the system rank ordering identical across the two datasets and two independent backbones.}

The rank ordering
$\text{M3-GDP}_{r_1}
 \!\succ\! \text{M3-GDP}_{r_0}
 \!\succ\! \text{M4}_{r_0}
 \!\approx\! \text{M3}_{r_0}
 \!\succ\! \text{M3}_{r_1}$
is identical across all three dataset--backbone combinations.
The illusion magnitude varies across conditions ($d{=}{-}0.08, {-}0.30, {-}1.32$) for two reasons.
First, D1's fixed 4-option format leaves less room for reasoning to diverge.
Second, M3~r1 undefined rates differ ($15.2\%, 14.2\%, 2.0\%$): a lower undefined rate leaves more hard cases in the sample, strengthening the measured illusion.

\paragraph{Is the gain just structural?}
Because \carahyb{} scores step-pair similarity, \gdp's structured \textsc{Claim}/\textsc{Ground}/\textsc{Stance} format could in principle inflate scores mechanically---through shared template vocabulary or cleaner step parsing---rather than through better reasoning alignment.
First, the format-only effect ($d{=}{+}0.62$) is less than half the full \gdp{} gain; the dominant component ($d{=}{+}1.51$) is the \textsc{Stance}-mediated debate interaction over content, not template.
Second, \cara-GT (Section~\ref{sec:results:caragt}) shows \gdp{} also aligns agents with the expert Scoring-Points reference ($d{=}{+}0.32$), which is not in \gdp{} format.
Third, GPT-4o judges M3-GDP~r1 reasoning as more aligned than M3~r1 on the same 50 agreement-set pairs ($4.86$ vs.\ $4.50$ on a 5-point scale, Cohen's $d{=}{+}0.75$, $95\%$ bootstrap CI on $\Delta$ $[{+}0.08, {+}0.65]$); since the LLM judge evaluates holistic logical coherence rather than template structure, this is direct evidence against pure structural confound.
Fourth, stripping the literal \textsc{Claim}/\textsc{Ground}/\textsc{Stance} tokens from \gdp{} traces and recomputing \carahyb{} attenuates paired Cohen's $d_z$ by only $6$--$10\%$ across all three conditions (Appendix~\ref{app:labelstripping}).

% ============================================================
% §8 Conclusion
% ============================================================
\section{Conclusion}
\label{sec:conclusion}

We identified the \emph{consistency illusion} in multi-agent medical QA: standard debate reduces detectable contradictions between agents while simultaneously decreasing the semantic similarity of their reasoning---agents appear to agree more but actually reason less consistently.
We introduced \cara, a family of metrics for cross-agent reasoning alignment, and applied it on MedQA-USMLE and MedThink-Bench: the illusion is negligible on D1 ($d{=}{-}0.08$) but significantly negative on D2 (Qwen $d{=}{-}0.30$; Llama-3 $d{=}{-}1.32$).
We proposed the Grounded Debate Protocol (\gdp), a prompt-level intervention requiring \textsc{Claim}+\textsc{Ground}+\textsc{Stance}, which delivers Tier-A alignment gains across all conditions ($d{=}{+}1.43$ to ${+}1.99$), eliminates two failure modes (complementary reasoning and sycophantic convergence) on both datasets and adds no LLM calls.
Together, our results challenge the assumption that multi-agent consensus indicates reasoning reliability, and show that protocol-level grounding can close this gap, positioning \gdp{} as a diagnostic and auditing intervention.
The structural sensitivity we observe---larger illusion where reasoning has more room to diverge---suggests grounding is most valuable on open-ended clinical tasks and proprietary backbones.

% ============================================================
% Limitations (after Conclusion per ACL 2023+ policy; does not count toward page limit)
% ============================================================
\section*{Limitations}
\label{sec:limitations}

While we evaluate \cara and \gdp across two medical QA benchmarks (D1 MedQA-USMLE $N{=}500$ and D2 MedThink-Bench $N{=}499$) and two open-weight backbones (Qwen 2.5 72B and Llama 3.3 70B), with Tier-A effect sizes ($d{=}{+}1.43$ to ${+}1.99$) replicated across all conditions, our analysis has several limitations.

\cara's agreement-set definition relies on discrete answer matching, which is well-defined for multiple-choice benchmarks but would require adaptation for free-form clinical generation tasks such as differential diagnosis or treatment-plan drafting; extending the metric to such settings is a natural next step.
Our cross-backbone validation covers two open-weight 70B-class models at comparable scales; proprietary models (e.g., GPT-4o, Claude, Gemini) would require an identical-instance experimental setup that is outside the scope of this work.

\carahyb is a step-level semantic-alignment measure designed to operate within the agreement set; the complementary GPT-4o oracle (Appendix~\ref{app:carallm}) and human-expert validation (Appendix~\ref{app:human_annotation}) anchor its interpretation, with both showing the predicted monotonicity across \carahyb terciles. Extending the metric to non-agreement-set settings (e.g., disagreement-pair audit) and to domains beyond clinical reasoning is left for future work.

\cara and \gdp target cross-agent reasoning alignment, a property orthogonal to answer accuracy; GDP is accordingly intended as an alignment-auditing and -improving intervention, and we make no claim that it improves accuracy. Across both backbones and datasets, majority-vote accuracy under GDP does not differ significantly from the standard debate baseline (McNemar tests; Table~\ref{tab:accuracy})---consistent with prior evidence that multi-agent debate itself does not reliably improve accuracy~\citep{choi2025martingale}. The Llama~3.3 comparison is based on $N{=}100$ and is underpowered to detect small accuracy effects; an adequately powered analysis of the alignment–accuracy relationship is left for future work.

We study a single, widely-used debate protocol---the symmetric, single-round design of \citet{du2024debate}, the canonical multi-agent debate baseline. Whether the consistency illusion and \gdp's correction generalize to other topologies (asymmetric roles, additional rounds, larger agent sets) is an important direction for future work.

% We therefore recommend reading a high CARA score as evidence of reasoning consistency rather than correctness, and reporting it alongside, not in place of, standard accuracy.

% ============================================================
% Ethics Statement (required for ACL/EMNLP)
% ============================================================
\section*{Ethics Statement}

This study uses publicly available medical QA benchmarks (MedQA-USMLE and MedThink-Bench) that contain no patient-identifiable information.
All LLM inference uses open-weight models (Qwen 2.5 72B, Llama 3.3 70B) running on institutional compute infrastructure.
\carallm validation uses the OpenAI API (GPT-4o) for 80 evaluation pairs.
No clinical participants are involved. Three medically-trained annotators rated 50 agent-pair outputs from the agreement set for metric validation; this is an annotation task on model outputs and does not constitute human-subjects research.

We emphasize that \cara measures reasoning \emph{alignment}, not reasoning \emph{correctness}.
High \cara scores indicate that agents reason consistently, not that they reason correctly.
\cara should be used as a diagnostic tool alongside accuracy metrics, not as a replacement for clinical validation.
Multi-agent medical QA systems evaluated in this work are research prototypes and are not intended for clinical deployment.

% Acknowledgments (empty for review)
% \section*{Acknowledgments}

% References
\bibliography{references}

% ============================================================
% Appendix
% ============================================================
\appendix

\section{Sample-Size Stability}
\label{app:stability}

A practical concern for any automated metric is whether scores are stable as the sample grows.
Table~\ref{tab:stability} compares earlier pilot \carahyb estimates (D1 $N{=}200$, D2 pooled $N{=}100$) against the full sample (D1 $N{=}500$, D2 $N{=}499$).
All ten cell--round--dataset combinations show deltas within $\pm 0.005$ (maximum: $0.005$ for M3~r0 on D2), with five of ten cells unchanged at the third decimal place.
This stability has two implications.
First, the metric is well-conditioned at the operational sample sizes used in the paper---further scaling would not change the qualitative conclusions.
Second, the pilot estimates were already reliable point estimates rather than artifacts of small-sample noise; the decision to scale up at $N{=}200$ accurately anticipated the final $N{=}500$ effect sizes.

\begin{table}[h]
\centering
\small
\setlength{\tabcolsep}{4pt}
\begin{tabular}{@{}llcccc@{}}
\toprule
& & \multicolumn{2}{c}{\textbf{D1}} & \multicolumn{2}{c}{\textbf{D2}} \\
\cmidrule(lr){3-4}\cmidrule(l){5-6}
\textbf{System} & \textbf{Round} & \textbf{Pilot} & \textbf{$\Delta$} & \textbf{Pilot} & \textbf{$\Delta$} \\
\midrule
M4     & r0 & 0.895 & $+0.000$ & 0.894 & $+0.001$ \\
M3     & r0 & 0.896 & $-0.001$ & 0.891 & $+0.005$ \\
M3     & r1 & 0.891 & $+0.000$ & 0.878 & $+0.003$ \\
M3-GDP & r0 & 0.914 & $-0.002$ & 0.913 & $-0.001$ \\
M3-GDP & r1 & 0.951 & $+0.001$ & 0.951 & $+0.001$ \\
\bottomrule
\end{tabular}
\caption{Pilot $\to$ full-sample \carahyb stability.
All deltas are within $\pm 0.005$ on both datasets, confirming that \carahyb is not sensitive to sample fluctuations at this scale.}
\label{tab:stability}
\end{table}

\section{Survivorship-Bias Sensitivity}
\label{app:sensitivity}

M3~r1 excludes the 76/500 (15.2\%) of D1 and 71/499 (14.2\%) of D2 questions where no majority answer emerged after debate.
These exclusions are non-random: harder questions are more likely to produce post-debate disagreement, so the surviving sample is biased toward inherently easier cases with higher baseline agreement.
The exclusion mechanism differs by system---M3-GDP retains $99.6\%$ of questions on both datasets---so direct comparison of M3 r1 against M3-GDP r1 (or against M4 r0, where M4 retains all D1 questions) involves comparing differently-selected populations.
We assess sensitivity through imputation: for the undefined questions, we impute plausible \carahyb{} and CR values and recompute the cell mean.
Three scenarios are reported in Table~\ref{tab:sensitivity}: \emph{observed} (current measurement, excluding undefined), \emph{neutral} (impute the M4 baseline value of that metric), and \emph{worst-case} (impute \carahyb${=}0.50$ and CR${=}1.0$---both far beyond any value observed in any cell on either dataset).

Two conclusions are robust across both datasets.
First, the \carahyb finding is robust: under all imputation scenarios---including the worst case, which assigns \carahyb${=}0.50$ (lower than any observed sample), M3~r1's adjusted score remains at or below M4~r0.
The consistency-illusion conclusion (M3 debate does not improve, and on D2 significantly degrades, alignment) therefore does not depend on the survivorship-biased sample.
Second, the CR finding is fragile in the opposite direction: under worst-case imputation, M3~r1's adjusted CR \emph{exceeds} M4~r0's on both datasets ($0.182{>}0.108$ on D1; $0.188{>}0.161$ on D2), so M3~r1's apparently low observed CR is at least partly an artifact of survivorship.
Practically, this means CR should be interpreted alongside the undefined rate, not in isolation, and the consistency-illusion claim rests primarily on the joint movement of CR and SIM in the main \carahyb result (Section~\ref{sec:results:main}) rather than on CR alone.

\begin{table}[t]
\centering
\small
\setlength{\tabcolsep}{4pt}
\begin{tabular}{@{}lcccc@{}}
\toprule
& \multicolumn{2}{c}{\textbf{D1 ($N{=}500$)}} & \multicolumn{2}{c}{\textbf{D2 ($N{=}499$)}} \\
\cmidrule(lr){2-3}\cmidrule(l){4-5}
\textbf{Scenario} & \textbf{HYB} & \textbf{CR} & \textbf{HYB} & \textbf{CR} \\
\midrule
M4 baseline       & 0.895 & 0.108 & 0.895 & 0.161 \\
\midrule
Observed (M3 r1)  & 0.892 & 0.035 & 0.881 & 0.054 \\
Neutral impute    & 0.892 & 0.046 & 0.883 & 0.069 \\
Worst-case        & \textbf{0.832} & \textbf{0.182} & \textbf{0.827} & \textbf{0.188} \\
\bottomrule
\end{tabular}
\caption{Sensitivity of M3~r1 to imputation of undefined questions.
\emph{Neutral} imputes the M4 baseline value; \emph{Worst-case} imputes HYB${=}0.50$ and CR${=}1.0$.
The HYB conclusion (M3~r1 $\leq$ M4) is robust on both datasets; the CR direction reverses under worst-case on both datasets.}
\label{tab:sensitivity}
\end{table}

\section{\cara-GT Per-Cell Results}
\label{app:caragt}

D2 includes expert-annotated reasoning steps (\emph{Scoring Points}) of $2$--$6$ steps per question (mean $3.1$), authored by clinical experts.
\cara-GT measures alignment between each agent's reasoning and the expert reference, computed via the same NLI$+$embedding pipeline used for \carahyb{} but with one agent replaced by the expert reference.
Where \carahyb{} asks whether agents reason consistently \emph{with each other}, \cara-GT asks whether they reason consistently \emph{with experts}.
\cara-GT is undefined when an agent produces zero extractable steps; this affects $35$/$2{,}495$ pairwise records ($1.4\%$), much cleaner than \carahyb's $156$/$2{,}495$ error rate (which additionally excludes agents who disagree on the final answer).
Table~\ref{tab:caragt} reports per-cell \cara-GT-HYB scores and expert-coverage rates on D2 ($N{=}499$).

\paragraph{\gdp improves expert alignment; standard debate degrades it.}
The pairwise comparisons mirror the inter-agent \carahyb{} pattern.
GDP$_{\text{r1}}$ vs.\ M3$_{\text{r1}}$: $\Delta\mathrm{GT\text{-}HYB}{=}{+}0.011$, $d{=}{+}0.32$, $95\%\,\mathrm{CI}{=}[{+}0.20,{+}0.45]$ (excludes zero).
M3$_{\text{r1}}$ vs.\ M4$_{\text{r0}}$: $\Delta{=}{-}0.007$, $d{=}{-}0.21$, $\mathrm{CI}{=}[{-}0.34,{-}0.08]$ (excludes zero).
The dual evidence---GDP improves alignment with respect to experts, and standard debate degrades it---supports the interpretation that the inter-agent illusion has a substantive directional component: under standard debate, agents drift not merely apart from each other but \emph{away} from the expert reasoning standard.

\paragraph{Expert coverage tells a sharper story.}
Coverage is the fraction of expert scoring points matched by at least one agent step at the contradiction-filtered alignment threshold.
Standard debate loses $8.6$ percentage points of expert coverage relative to independent voting (M4: $88.8\%$ $\to$ M3~r1: $80.2\%$, $\mathrm{CI}{=}[{-}11.2,{-}6.1]$): roughly one in eleven expert steps is dropped during debate.
\gdp debate recovers $5.4$pp of that loss (M3~r1: $80.2\%$ $\to$ M3-GDP~r1: $85.6\%$, $\mathrm{CI}{=}[{+}2.7,{+}8.1]$).
Standard debate \emph{drops} expert scoring points; \gdp debate \emph{keeps} them.
This coverage gap is independent of the inter-agent \carahyb{} measurement and provides a second, expert-grounded line of evidence for the same conclusion: \gdp produces reasoning that is more consistent with both other agents and external standards.

\begin{table}[t]
\centering
\small
\setlength{\tabcolsep}{4pt}
\begin{tabular}{@{}llccc@{}}
\toprule
\textbf{System} & \textbf{Round} & \textbf{GT-HYB} & \textbf{Coverage} & \textbf{$N$} \\
\midrule
M3-GDP & r1 & \textbf{0.813} & \textbf{85.6\%} & 499 \\
M3     & r0 & 0.810 & 87.4\% & 499 \\
M4     & r0 & 0.809 & 88.8\% & 499 \\
M3-GDP & r0 & 0.807 & 88.2\% & 499 \\
M3     & r1 & 0.802 & 80.2\% & 464 \\
\bottomrule
\end{tabular}
\caption{\cara-GT on D2 ($N{=}499$): alignment between agent reasoning and expert \emph{Scoring Points}.
Coverage is the fraction of expert scoring points that match at least one agent step.}
\label{tab:caragt}
\end{table}

\section{Accuracy and McNemar Tests}
\label{app:accuracy}

\cara{} measures reasoning alignment, not answer correctness, but we separately report majority-vote accuracy for context.
Accuracy is computed by extracting the majority answer per agent at each round, then comparing to the gold answer with continuity-corrected McNemar tests on paired binary outcomes (correct/incorrect per question) and paired percentile bootstrap CIs ($B{=}10{,}000$, seed $42$) on the accuracy delta.
Table~\ref{tab:accuracy} reports majority-vote accuracy for all dataset--backbone combinations.

\paragraph{Findings.}
Standard debate (M3 r1) does not significantly change accuracy on either Qwen dataset (D1: $\Delta{=}{+}0.4$pp, $p{=}0.86$; D2: $\Delta{=}0.0$pp, $p{=}0.90$), consistent with the martingale characterization of debate dynamics: gains from sampling and voting are already realized by the independent baseline.
D2 baseline accuracy ($\sim 41\%$) is much lower than D1 ($\sim 84\%$) because D2 questions have a mean of $\sim 7.5$ answer options (random baseline $\sim 13\%$), making $\sim 41\%$ approximately $3{\times}$ random on this benchmark---i.e., the model is performing non-trivially despite the lower headline number.
Llama D2 accuracy ($N{=}100$, $49.0\%$) and Qwen D2 ($N{=}499$, $41.5\%$) are computed on different sample sizes of the same dataset, so we do not draw cross-backbone accuracy conclusions from this comparison.

\begin{table*}[t]
\centering
\small
\setlength{\tabcolsep}{4pt}
\begin{tabular}{@{}lccccc@{}}
\toprule
\textbf{Dataset} & \textbf{Backbone} & \textbf{M4} & \textbf{M3 r1} & \textbf{GDP r1} & \textbf{$\Delta_{\text{GDP}}$} \\
\midrule
D1 MedQA ($N{=}500$)    & Qwen  & 83.8\% & 84.2\% & 81.4\% & $-2.4$ pp$^{\text{ns}}$ \\
D2 MedThink ($N{=}499$) & Qwen  & 41.5\% & 41.5\% & 39.1\% & $-2.4$ pp$^{\text{ns}}$ \\
D2 MedThink ($N{=}100$) & Llama & 49.0\% & 47.0\% & 43.0\% & $-6.0$ pp$^{\text{ns}}$ \\
\bottomrule
\end{tabular}
\caption{Majority-vote accuracy.
$\Delta_{\text{GDP}}{=}$ GDP$_{\text{r1}}{-}$M4$_{\text{r0}}$.
``ns'' = not significant at $\alpha{=}0.05$.
On D1, GDP vs.\ M3: McNemar $p{=}0.061$; GDP vs.\ M4: $p{=}0.097$; M3 vs.\ M4: $p{=}0.86$.
On D2 (Qwen), all pairwise McNemar tests yield $p{>}0.15$.
On D2 (Llama, $N{=}100$), all pairwise tests yield $p{>}0.28$.}
\label{tab:accuracy}
\end{table*}

\section{\carahyb Implementation}
\label{app:carahyb_impl}

\paragraph{NLI component.}
DeBERTa-v3-large fine-tuned on MNLI, FEVER, ANLI, LingNLI, and WANLI~\citep{laurer-etal-2024-less} (304M parameters).
We threshold the contradiction probability at $\tau{=}0.7$ for the hybrid filter.

\paragraph{Embedding component.}
Stella-EN-1.5B-v5~\citep{zhang2024stella} (1.5B parameters), L2-normalized $1024$-dimensional vectors; cosine similarity reduces to dot product.

\paragraph{Step decomposition.}
Deterministic regex pipeline: primary extraction of numbered lists (e.g., \texttt{1.}, \texttt{2)}) and bullet points (\texttt{-}, \texttt{*}); sentence-level splitting as fallback for unstructured paragraph responses.
The final answer line (e.g., \texttt{ANSWER: X}) is stripped before decomposition; resulting steps shorter than 20 characters are filtered.

\paragraph{Runtime.}
On a single NVIDIA RTX 4090 (24~GB), \cara processes ${\sim}9{,}000$ traces per hour.
The full D1 pipeline ($N{=}500$, $2{,}500$ pairwise records) completes in ${\sim}8$ minutes; D2 ($N{=}499$, $2{,}495$ records) in ${\sim}8$ minutes.

\section{\carallm Oracle Validation}
\label{app:carallm}

\paragraph{Design.}
To validate \carahyb against human-like judgment, we conduct a three-dimensional validation using GPT-4o~\citep{openai2024gpt4o} as an alignment judge.
We evaluate $80$ agent pairs: $50$ agreement-set pairs (both agents selected the majority answer) and $30$ disagreement pairs (agents selected different answers).
The inclusion of disagreement pairs is critical: agreement-set pairs produce a ceiling effect in GPT-4o scores (skewed toward $4$--$5$), and disagreement pairs break this ceiling by introducing genuinely misaligned reasoning (scores $2$--$3$), expanding the effective scoring range.
For each pair, GPT-4o assigns a $1$--$5$ alignment score at temperature $0.0$, repeated three times per pair. We report \carallm{} as this rating normalized to $[0,1]$ via $(\text{score}{-}1)/4$; the per-system head-to-head below is given on the raw $1$--$5$ scale for interpretability.

\paragraph{Three-dimensional validation results.}
\begin{enumerate}[nosep,leftmargin=*]
\item \textbf{Moderate pair-level correlation} (Spearman $\rho{=}0.55$, $n{=}80$): \carahyb and GPT-4o capture overlapping but non-identical dimensions of alignment.
The correlation is moderate rather than high because the two metrics operationalize alignment differently: \carahyb measures step-level semantic overlap via embeddings, while GPT-4o evaluates holistic logical coherence.
\item \textbf{Strict stratum monotonicity}: when pairs are binned by \carahyb terciles, mean \carallm scores are $0.67$ (low), $0.77$ (mid), $0.82$ (high).
\carahyb reliably distinguishes alignment levels even where it does not predict individual GPT-4o scores precisely.
\item \textbf{Perfect inter-run reliability}: all $80$ pairs receive identical GPT-4o scores across three independent runs (inter-run SD $= 0.00$), confirming deterministic evaluation at temperature~$0$.
\end{enumerate}

\paragraph{Per-system head-to-head.}
Within the $50$ agreement-set pairs, M3-GDP~r1 receives a mean GPT-4o score of $4.86$ ($n{=}28$, SD $0.35$), vs.\ $4.50$ for M3~r1 ($n{=}22$, SD $0.58$); Cohen's $d{=}{+}0.75$, $95\%$ bootstrap CI on $\Delta$ $[{+}0.08, {+}0.65]$ excludes zero ($B{=}10{,}000$, seed $42$).
This direct head-to-head corroborates the inter-agent \carahyb{} ranking and complements the tercile monotonicity reported above.

\paragraph{Disagreement-pair divergence.}
On disagreement pairs (agents with different final answers), \carahyb assigns scores of $0.83$--$0.95$ because medical reasoning steps have high embedding similarity even when conclusions differ, while GPT-4o assigns $2$--$3$ out of $5$ because it detects the logical incoherence of conflicting conclusions.
This divergence is by design: \carahyb measures reasoning-step alignment \emph{within the agreement set}, so its high scores on disagreement pairs reflect genuine semantic overlap in medical reasoning vocabulary rather than a metric failure.
The complementary roles ground the joint use of \carahyb (scalable, surface alignment) and \carallm (oracle, logical coherence) we recommend in deployment.

\section{Label-Stripped \carahyb: Structural Confound Check}
\label{app:labelstripping}

\paragraph{Motivation.}
\carahyb{} computes step-pair embedding cosine and NLI contradictions over reasoning steps extracted from agent responses.
\gdp{} requires each step to carry literal \textsc{Claim}/\textsc{Ground}/\textsc{Stance} label tokens, so a reasonable concern is that these template tokens themselves inflate the embedding component of \carahyb{}, making part of the \gdp-vs-M3 gap a metric artifact of template self-similarity rather than substantive reasoning convergence.
The format/debate decomposition (Section~\ref{sec:analysis}) and the \cara-GT result (Section~\ref{sec:results:caragt}) already weigh against this concern; this appendix reports the direct check.

\paragraph{Procedure.}
For each \gdp{} trace (D1 Qwen, D2 Qwen, D2 Llama; both $r_0$ and $r_1$) we remove the literal label tokens \texttt{CLAIM:}, \texttt{GROUND:}, and \texttt{STANCE:} together with their Markdown bold variants (e.g., \texttt{**CLAIM**}), preserving (i) the numbered-step markers (\texttt{1.}, \texttt{2.}, \ldots) that M3 traces also use, (ii) the \textsc{Agree}/\textsc{Disagree}/\textsc{Extend} content tokens that constitute the agent's actual stance decision, and (iii) all free-text claim, ground, and stance content.
We then re-run the unchanged \carahyb{} pipeline (DeBERTa-v3 NLI + Stella embeddings, $\tau{=}0.7$, seed $42$) on the stripped traces; M3 and M4 cells are unchanged.
Paired Cohen's $d_z$ is computed at the question level for \gdp{}~r1 vs.\ M3~r1, using the same script for both the original and the stripped condition.

\paragraph{Results.}
Table~\ref{tab:labelstripping} reports the paired comparison across all three dataset-backbone conditions.
Removing the literal label tokens attenuates Cohen's $d_z$ by only $6$--$10\%$.
All three $95\%$ bootstrap CI lower bounds remain above $+0.76$, and mean \carahyb{} itself drops by only ${\sim}0.005$ across configurations.
Approximately $90\%$ of the alignment effect survives label removal, indicating that the literal template tokens are not the principal source of \gdp's gain on \carahyb{}.

\begin{table*}[h]
\centering
\small
\setlength{\tabcolsep}{4pt}
\begin{tabular}{@{}lccc@{}}
\toprule
\textbf{Config} & \textbf{Original $d_z$} & \textbf{Stripped $d_z$} & \textbf{Stripped 95\% CI} \\
\midrule
D1 MedQA (Qwen, $n{=}422$)    & $+0.97$ & $+0.87$ & $[+0.76, +0.99]$ \\
D2 MedThink (Qwen, $n{=}427$) & $+1.16$ & $+1.09$ & $[+0.98, +1.21]$ \\
D2 MedThink (Llama, $n{=}98$) & $+1.39$ & $+1.29$ & $[+0.96, +1.74]$ \\
\bottomrule
\end{tabular}
\caption{Label-stripped \carahyb{} confound check.
Paired Cohen's $d_z$ between \gdp~r1 and M3~r1 on the agreement set, before and after removing literal \textsc{Claim}/\textsc{Ground}/\textsc{Stance} tokens from \gdp{} traces.
The effect attenuates by $6$--$10\%$ but remains Tier~B--A in all three conditions, with all CI lower bounds above $+0.76$.}
\label{tab:labelstripping}
\end{table*}

\paragraph{Interpretation and limitations.}
The label-stripping experiment isolates one specific structural-confound mechanism---that the literal template tokens themselves drive embedding similarity.
The data do not support this mechanism: ${\sim}90\%$ of the effect persists after token removal.

\paragraph{Robustness to the NLI model.}\carahyb{} is also robust to the choice of contradiction detector: substituting a cross-encoder NLI model (\texttt{cross-encoder/nli-deberta-v3-large}) for the default DeBERTa-MNLI on the D2 $N{=}100$ set leaves the illusion direction unchanged (M3~r1 ${<}$ M4~r0) and Cohen's $d$ nearly identical ($-0.36$ vs.\ $-0.38$).

\section{Human-Expert Validation}
\label{app:human_annotation}

\paragraph{Setup.}
Three independent annotators with medical training each rated $50$ agent pairs drawn from the agreement set on the MedThink-Bench (D2) benchmark.
The 50 pairs are stratified by \carahyb tercile (low/mid/high, $\sim 17$ pairs each).
Each pair was rated on a 5-point alignment scale ($5{=}$near-identical reasoning, $4{=}$substantially aligned, $3{=}$partially aligned, $2{=}$weakly aligned, $1{=}$misaligned reasoning with the same final answer) and a 3-point confidence scale, following the annotation guidelines released with the code repository.
Annotators saw only the question and the two agents' reasoning steps; system identity (M3 vs.\ M3-GDP) and \carahyb scores were withheld to prevent bias.

\paragraph{Inter-annotator agreement.}
Pairwise Spearman correlation between annotators is consistently high (Table~\ref{tab:iaa}); $100\%$ of pair-level disagreements are bounded to $\leq 1$ scale point, confirming task reliability.

\begin{table}[h]
\centering
\footnotesize
\setlength{\tabcolsep}{3.5pt}

\begin{tabular}{@{}lccc@{}}
\toprule
\textbf{Annotator pair} &
\makecell[c]{\textbf{Spearman} \\ \textbf{$\rho$}} &
\makecell[c]{\textbf{Exact} \\ \textbf{agreement}} &
\makecell[c]{\textbf{Within-1}} \\
\midrule
A1 vs.\ A2 & 0.934 & 76\% & 100\% \\
A1 vs.\ A3 & 0.987 & 98\% & 100\% \\
A2 vs.\ A3 & 0.927 & 74\% & 100\% \\
\bottomrule
\end{tabular}

\caption{Inter-annotator agreement on the 50-pair human annotation.}
\label{tab:iaa}
\end{table}

\paragraph{Correlation with CARA metrics.}
Aggregate human ratings (mean of three annotators) correlate positively with \carahyb (Spearman $\rho{=}0.263$), \carasim ($\rho{=}0.312$), and weakly with the contradiction rate ($\rho{=}0.137$).
Each individual annotator shows a similar correlation with \carahyb ($\rho{=}0.21$--$0.26$), confirming the signal is not driven by any single annotator.

\begin{figure}[t]
\centering
\includegraphics[width=\columnwidth]{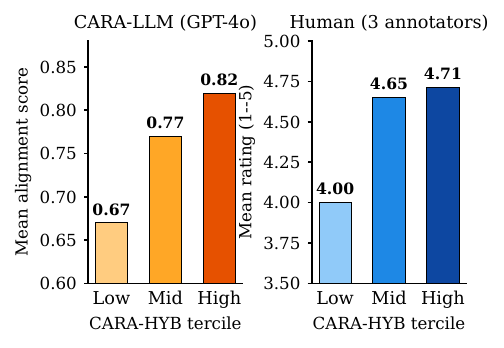}
\caption{Tercile monotonicity across two external validators on the same 50-pair sample stratified by \carahyb. Both the GPT-4o oracle (\carallm; Appendix~\ref{app:carallm}) and the mean of three medically-trained annotators show monotonic increase from low to mid to high \carahyb strata.}
\label{fig:tercile_monotonicity}
\end{figure}

\paragraph{Tercile monotonicity.}
Aggregate ratings increase monotonically across \carahyb strata: $4.00$ (low, $n{=}17$), $4.65$ (mid, $n{=}17$), $4.71$ (high, $n{=}16$).
The same monotonic pattern is observed in the \carallm GPT-4o oracle terciles (Figure~\ref{fig:tercile_monotonicity}), supporting a consistent ordering of \carahyb-graded alignment across two independent external judges.
The low-to-mid gap is larger than the mid-to-high gap, reflecting a real shift in alignment quality at the lower end combined with a ceiling effect at the upper end (more than half of high-stratum pairs receive the maximum rating of $5$).

\paragraph{Range restriction.}
All 50 pairs have $\carahyb \in [0.76, 0.98]$ because the metric is conditioned on the agreement set (Section~\ref{sec:cara:setup}); the ``low'' stratum mean is $0.864$, far above the corpus-level low end. This narrow range bounds the achievable Spearman correlation against any external judge, paralleling the bounded correlation observed in the \carallm GPT-4o oracle validation when restricted to the same agreement set.

\paragraph{Annotation instructions.}
Annotators received the following written guidelines before the task (verbatim text is released with our anonymized code repository):

\begin{tcolorbox}[colback=gray!6, colframe=gray!50, boxrule=0.4pt, arc=2pt,
left=4pt, right=4pt, top=3pt, bottom=3pt, fontupper=\small]
\textbf{Task overview.} You will see pairs of AI-generated medical
reasoning traces. Both agents chose the same final answer; judge how
aligned their reasoning is---not whether the answer is correct.

\smallskip
\textbf{Rating 1 --- Reasoning Alignment (1--5):}
\begin{itemize}[nosep, leftmargin=*]
\item \textbf{5 (Near-identical):} same claims, same facts, same logical sequence; minor wording differences only.
\item \textbf{4 (Substantially aligned):} same core reasoning and key facts; one may add detail or reorganize.
\item \textbf{3 (Partially aligned):} share key steps but diverge on intermediate justifications.
\item \textbf{2 (Weakly aligned):} same answer through largely different reasoning; few overlapping steps.
\item \textbf{1 (Misaligned):} same answer through fundamentally different or contradictory reasoning.
\end{itemize}

\smallskip
\textbf{Rating 2 --- Confidence (1--3):} $3$~confident, $2$~somewhat confident, $1$~low confidence (reasoning too vague to
judge).

\smallskip
\textbf{What NOT to judge.} Do not evaluate answer correctness, writing quality, response format (structured vs.\ free
text), or step count. Some responses use a \textsc{Claim}/\textsc{Ground}/\textsc{Stance} format; others use free
text---judge reasoning content regardless of format. System identity and automated \cara{} scores are withheld to prevent
bias.
\end{tcolorbox}

\paragraph{Annotator background and procedure.}
The three annotators are medically-trained professionals aged $30$--$35$ (two female resident physicians and one male
postdoctoral fellow), recruited through direct personal contact by the authors and participating as volunteers without
compensation. All three annotators rated all $50$ pairs, enabling the pairwise Spearman inter-annotator agreement reported
above. Pairs were presented in randomized order, and system identity (M3 vs.\ M3-GDP), agent round, and
\carahyb{}/\carasim{} scores were withheld throughout. The task does not constitute human-subjects research---annotators
rated machine-generated outputs only---and required no institutional review. The verbatim annotation guidelines, blank annotation sheet, and three completed per-annotator scored sheets are released with our anonymized code repository.

\section{\cara Variants and Aggregation Formulas}
\label{app:variants}

\paragraph{Variant table.}
The four \cara alignment variants used in the paper are summarized below.

\begin{table*}[h]
\centering
\small
\begin{tabular}{@{}llp{9cm}@{}}
\toprule
\textbf{Variant} & \textbf{Range} & \textbf{Definition} \\
\midrule
\caranli & $\{-1, 0, +1\}$ & NLI label: $+1$ entailment, $0$ neutral, $-1$ contradiction. \\[3pt]
\carasim & $[-1, +1]$ & Cosine similarity: $\cos(\mathrm{emb}(r_i^k), \mathrm{emb}(r_j^l))$. \\[3pt]
\carahyb & $[-1, +1]$ & Hybrid: if $P(\text{contradiction}){>}\tau$ then $-1$, else cosine similarity. \\[3pt]
\carallm & $[0, 1]$ & GPT-4o $1$--$5$ alignment rating, normalized to $[0,1]$ via $(\text{score}{-}1)/4$. \\
\bottomrule
\end{tabular}
\caption{\cara alignment variants. \carahyb is the default metric used throughout; \carallm serves as a validation oracle.}
\label{tab:cara-variants}
\end{table*}

\paragraph{Rescaling.}
For variants whose range includes negative values (\caranli, \carahyb), we define a unit-interval rescaling:
\begin{equation}
  \mathrm{CARA}_{[0,1]} = \frac{\mathrm{CARA} + 1}{2}.
  \label{eq:rescale}
\end{equation}

\paragraph{Contradiction Rate.}
The Contradiction Rate diagnostic is defined in Eq.~(\ref{eq:cr}) of Section~\ref{sec:cara:aggregation}.
The denominator $\sum_{i \in \mathcal{S}(q)}(|\mathcal{S}(q)|{-}1) K_i$ equals $\sum_{i<j} (K_i + K_j)$, the total number of best-match operations across all directed agent pairs in the agreement set.
CR captures the rate at which best-aligned step pairs across agreeing agents are nonetheless contradictory: a direct indicator of the consistency illusion.

\section{System Configurations}
\label{app:systems}

\paragraph{Configuration overview.}
We evaluate three system configurations under identical compute budgets.
Each system uses three agents and aggregates their final answers by majority vote.
``Round~0'' refers to the independent step in which each agent answers without seeing the others' responses; ``Round~1'' refers to the post-debate step in which each agent revises its answer after seeing all other agents' Round~0 outputs.
M4 has only Round~0 (the control condition with no debate); M3 and M3-GDP have both rounds.
Majority voting is applied to Round~1 answers for M3 and M3-GDP, and to Round~0 answers for M4.

\begin{table}[h]
\centering
\small
\setlength{\tabcolsep}{4pt}

\begin{tabular}{@{}lp{2.7cm}ccc@{}}
\toprule
\textbf{ID} & \textbf{System} & \textbf{Agents} & \textbf{Rounds} & \textbf{GDP} \\
\midrule
M3 & \makecell[l]{\citet{du2024debate}\\ Debate} & 3 & 1 & no \\
M4     & Independent Vote            & 3 & 0 & no \\
M3-GDP & Debate $+$ \gdp             & 3 & 1 & yes \\
\bottomrule
\end{tabular}

\caption{System configurations.
All systems use majority voting for answer aggregation.
``Rounds'' counts debate rounds; Round~0 (independent) is always present.
M4 serves as a no-communication control.}

\label{tab:systems}
\end{table}

\paragraph{Backbone serving environment.}
Primary trials use Qwen~2.5 72B Instruct AWQ (4-bit quantized) served via vLLM~\citep{kwon2023vllm} on $2{\times}$NVIDIA A40 GPUs (48~GB each).
The cross-backbone replication uses Llama~3.3 70B Instruct AWQ on a separate pair of A40 GPUs with the same vLLM serving configuration.
All experiments use the same decoding hyperparameters: temperature $0.7$, top-$p$ $0.9$, and a maximum of $2{,}048$ output tokens per agent response.

\paragraph{Standard M3 and M4 prompts.}
M3 and M4 share a 2-message Round~0 prompt.
The system message is verbatim:
\begin{quote}\small
\texttt{You are a medical specialist. Answer medical questions with step-by-step reasoning.}
\end{quote}
The user message contains the question text with inline answer options, followed by the suffix:
\begin{quote}\small
\texttt{Think step by step and provide your answer in the format: ANSWER: X}
\end{quote}
M3 Round~1 extends this to a 4-message asymmetric prompt: the original system and user messages, the agent's own Round~0 response as the assistant turn, and a final user turn presenting the other agents' Round~0 responses with the instruction \emph{``Using these as additional information, provide your answer in the format: ANSWER: X.''}
Each agent sees only the other two agents' responses in the debate user turn (never its own), preventing self-conditioning artifacts.

\paragraph{M3-GDP prompts.}
M3-GDP preserves the 2-message (Round~0) and 4-message (Round~1) prompt structure of M3 but replaces the system message with format requirements that enforce structured reasoning.
The Round~0 system message requires \textsc{Claim}+\textsc{Ground} for each reasoning step, where \textsc{Ground} must name a specific medical fact, mechanism, or guideline supporting the claim.
The Round~1 system message adds \textsc{Stance} (\textsc{Agree}/\textsc{Disagree}/\textsc{Extend}) with respect to each other agent's claims, with the rule that any \textsc{Disagree} must be accompanied by a counter-\textsc{Ground}.
Round~1 also includes an explicit anti-sycophancy clause:
\begin{quote}\small
\texttt{Do NOT change your answer simply because other agents disagree. Only change if you identify a specific factual error in your own reasoning or receive a more compelling medical ground from another agent.}
\end{quote}
The full GDP Round~0 and Round~1 system messages, along with the debate user template, are released with the code repository.

\section{GDP Worked Example}
\label{app:gdp_example}

We compare a standard debate (M3) trace and a \gdp{} (M3-GDP) trace on the same clinical question.
Both systems converge on the same correct answer; the contrast is in how the agents arrive there.

\begin{tcolorbox}[colback=gray!8, colframe=gray!50, boxrule=0.4pt, arc=2pt, left=4pt, right=4pt, top=3pt, bottom=3pt]
\small
\textbf{Clinical question.}~A patient presents with fasting plasma glucose of $140$~mg/dL on two consecutive measurements.
What is the most appropriate first-line management?
\begin{itemize}[nosep, leftmargin=*, label={}]
\item[] A.~Insulin \quad \textbf{B.~Metformin} \quad C.~Lifestyle modification only \quad D.~Sulfonylurea
\end{itemize}
Gold answer: \textbf{B}.
\end{tcolorbox}

\vspace{4pt}
\noindent\textbf{M3 (Standard Debate) --- Round~1}

\begin{tcolorbox}[colback=red!4, colframe=red!40, boxrule=0.4pt, arc=2pt, title={\small Agent~A}, fonttitle=\small\bfseries, left=4pt, right=4pt, top=2pt, bottom=2pt]
\small
The elevated fasting glucose is consistent with a diagnosis of diabetes.
Considering the clinical picture, metformin would be appropriate.
I agree with Agent~B's reasoning.
\textit{Answer: B.}
\end{tcolorbox}

\begin{tcolorbox}[colback=red!4, colframe=red!40, boxrule=0.4pt, arc=2pt, title={\small Agent~B}, fonttitle=\small\bfseries, left=4pt, right=4pt, top=2pt, bottom=2pt]
\small
The patient's clinical picture supports metformin therapy.
Given the lab values and the analysis Agent~A provided, I support this approach.
\textit{Answer: B.}
\end{tcolorbox}

\noindent\textit{\footnotesize Voted solution: \textbf{B}~\checkmark{}~~|~~Both agents converge with no medical citations; agreement is vague.}

\vspace{6pt}
\noindent\textbf{M3-GDP --- Round~1}

\begin{tcolorbox}[colback=green!4, colframe=green!40, boxrule=0.4pt, arc=2pt, title={\small Agent~A}, fonttitle=\small\bfseries, left=4pt, right=4pt, top=2pt, bottom=2pt]
\small
\texttt{CLAIM:} The fasting glucose of $140$~mg/dL meets ADA diagnostic criteria for T2DM. \\
\texttt{GROUND:} Fasting plasma glucose ${\geq}126$~mg/dL on two occasions confirms diabetes (ADA Standards of Care).\\
\texttt{STANCE:} \textsc{Disagree} with Agent~B (insulin)---insulin is not first-line for uncomplicated T2DM without DKA.
\textit{Answer: B.}
\end{tcolorbox}

\begin{tcolorbox}[colback=green!4, colframe=green!40, boxrule=0.4pt, arc=2pt, title={\small Agent~B}, fonttitle=\small\bfseries, left=4pt, right=4pt, top=2pt, bottom=2pt]
\small
\texttt{CLAIM:} Metformin is the appropriate first-line agent for newly diagnosed T2DM. \\
\texttt{GROUND:} ADA Standards of Care recommend metformin first-line for adults with T2DM without contraindications, given its efficacy in reducing hepatic glucose production and favorable safety profile.\\
\texttt{STANCE:} \textsc{Agree} with Agent~A's diagnostic claim; reaffirm metformin over insulin given absence of DKA or severe hyperglycemia.
\textit{Answer: B.}
\end{tcolorbox}

\noindent\textit{\footnotesize Voted solution: \textbf{B}~\checkmark{}~~|~~Both agents cite ADA criteria and take explicit stances on each other's claims; agreement is grounded.}

\vspace{6pt}
\noindent
The contrast is structural: M3 agents agree on the answer without committing to specific medical evidence or explicitly engaging with each other's claims, whereas M3-GDP agents reach the same answer through named criteria (ADA guidelines) and explicit \textsc{Stance} declarations.
This pattern recurs at the dataset scale, producing the \carahyb{} and \cara-GT gaps reported in Section~\ref{sec:results:main} and Appendix~\ref{app:caragt}.

\section{Failure Mode Taxonomy and Case Study}
\label{app:failure_modes}

\paragraph{Failure-mode codebook.}
The 50 lowest-\carahyb correct-answer cases per system are classified using deterministic rules over CARA sub-metrics (CR${>}0.3$, step-count ratio${>}2$, zero-step agents, SIM${<}0.80$).

\begin{table}[h]
\centering
\small
\setlength{\tabcolsep}{2pt}
\begin{tabular}{@{}lp{0.62\columnwidth}@{}}
\toprule
\textbf{Code} & \textbf{Failure mode} \\
\midrule
FM1 & \textsc{Complementary Reasoning}: different but non-contradictory reasoning paths. Low SIM, low CR. \\
FM2 & \textsc{Granularity Mismatch}: step-count ratio ${>}2$. Depth difference, not substantive disagreement. \\
FM3 & \textsc{Contradictory Premises}: agents cite contradictory facts but converge on the correct answer. \\
FM4 & \textsc{Sycophantic Convergence}: agent adopts the majority answer without substantive reasoning (zero steps). \\
FM5 & \textsc{Terminology Divergence}: different medical terms for the same concept. \\
FM6 & \textsc{Partial Overlap}: residual baseline misalignment without extreme signals. \\
\bottomrule
\end{tabular}
\caption{Six failure modes identified by classifying the 50 lowest-\carahyb correct-answer cases per system.}
\label{tab:fm_taxonomy}
\end{table}

\paragraph{Case study: MQA\_0447 (hypovolemic shock).}
We illustrate FM3 (contradictory premises) with a D1 case in which the correct answer is reached through partially incorrect reasoning.
The question describes a $52$-year-old man with variceal bleeding and hypovolemic shock.
Two of three \gdp agents in the agreement set initially state that systemic vascular resistance (SVR) \emph{decreases} in hypovolemic shock---a clinically incorrect claim, since SVR \emph{increases} as a compensatory response to maintain perfusion.
Both agents then self-correct within the same response and select the correct treatment (volume resuscitation followed by endoscopic intervention).
This case exemplifies the consistency illusion at the individual-fact level: the answer is correct, but the underlying reasoning contains explicit factual contradictions that the metric correctly surfaces.
\gdp's structured format makes these contradictions \emph{detectable}; under standard debate, the same misstatement would more likely be smoothed away in vague free text.

\section{Undefined Rate}
\label{app:undefined}

The undefined rate---questions where no majority answer emerges---is an independent behavioral indicator that does not depend on agreement-set selection.
Table~\ref{tab:undefined} reports per-system rates on both datasets.

\paragraph{Standard debate destabilizes consensus.}
On D1, M4 (independent voting) achieves $100\%$ consensus because agents sample independently from the same distribution: the probability of three-way disagreement among i.i.d.\ samples is low.
Standard debate (M3 r1) drives the undefined rate to $15.2\%$ on D1---a $38{\times}$ increase over the pre-debate round (M3~r0, $0.4\%$).
On D2, where the harder question distribution gives even M4 a $5.8\%$ undefined rate, M3 r1 still pushes this up to $14.2\%$ ($2.5{\times}$ relative).
\gdp debate maintains consensus at near-M4 levels ($0.4\%$ on both datasets) while substantially improving reasoning alignment.

\paragraph{Reasoning collapse.}
A related phenomenon is reasoning collapse: agents that produce zero extractable reasoning steps after debate (a behavioral signature of sycophantic convergence).
On D2 N=100, M3 r1 yields zero-step agents in $5.0\%$ of agreement-set memberships; M3-GDP r1 in only $0.3\%$.
\gdp's \textsc{Claim}+\textsc{Ground} requirement prevents this collapse because an agent cannot adopt an answer without producing a named ground supporting it.

\begin{table}[t]
\centering
\small
\setlength{\tabcolsep}{6pt}
\begin{tabular}{@{}lcc@{}}
\toprule
\textbf{System / Round} & \textbf{D1 ($N{=}500$)} & \textbf{D2 ($N{=}499$)} \\
\midrule
M3 r1     & 76 (15.2\%) & 71 (14.2\%) \\
M3-GDP r0 & 3  (0.6\%)  & 26 (5.2\%) \\
M3 r0     & 2  (0.4\%)  & 28 (5.6\%) \\
M3-GDP r1 & 2  (0.4\%)  & 2  (0.4\%) \\
M4 r0     & 0  (0.0\%)  & 29 (5.8\%) \\
\bottomrule
\end{tabular}
\caption{Undefined rate per system--round--dataset.
M3 r1 destabilizes consensus on ${\sim}14$--$15\%$ of questions on both datasets; M3-GDP r1 retains essentially all of them.}
\label{tab:undefined}
\end{table}

\section{Consistency-Illusion Visualization}
\label{app:illusion_2d}

\begin{figure}[h]
\centering
\includegraphics[width=\columnwidth]{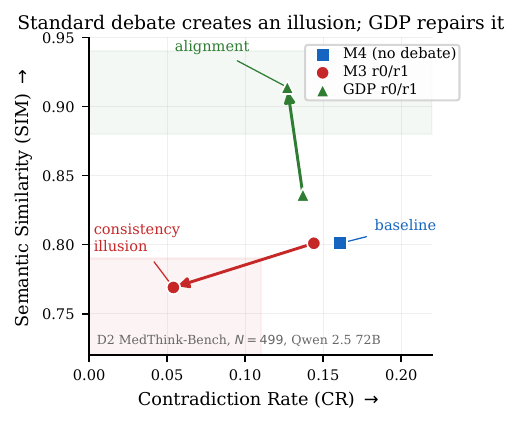}
\caption{Empirical (CR, SIM) trajectory on D2 ($N{=}499$, Qwen 2.5 72B).
M3 (red arrow) moves r0$\,\to\,$r1 \emph{both} CR$\,\downarrow$ \emph{and} SIM$\,\downarrow$ (consistency illusion).
\gdp (green arrow) moves r0$\,\to\,$r1 SIM$\,\uparrow$ while keeping CR informative.}
\label{fig:illusion_2d}
\end{figure}

Figure~\ref{fig:illusion_2d} visualizes the (CR, SIM) trajectory of each system on D2 ($N{=}499$, Qwen 2.5 72B), making the dual movement of standard debate (CR$\,\downarrow$ + SIM$\,\downarrow$) and the repair pattern under \gdp (SIM$\,\uparrow$) directly observable.
The two arrows correspond to the r0$\,\to\,$r1 transitions under M3 (red) and M3-GDP (green); M4 is shown as a single point for reference.
The figure complements the conceptual schematic of Figure~\ref{fig:illusion_concept} with the actual measured coordinates that anchor every claim made in Section~\ref{sec:results:main}.

\section{Mechanism Discussion}
\label{app:mechanism_extended}

The condensed mechanism account in RQ1 (Section~\ref{sec:analysis}) merges two effects under one heading.
Here we elaborate each with the original four-component analysis.

\paragraph{Contradiction smoothing without reasoning convergence (M3).}
During debate, agents see each other's responses and revise their own.
When agents disagree on factual claims, revision pressure causes them to soften or remove the conflicting statements, which reduces CR.
However, this removal is \emph{subtraction without alignment}: agents delete contradictory steps without replacing them with shared reasoning anchors that other agents have also committed to.
The result is fewer contradictions but also less overlapping reasoning content, so SIM decreases.
This pattern is stronger on D2 because the more open answer space (3--10 options, mean ${\sim}7.5$) gives agents more room to diverge in reasoning while still landing on the same final answer.

\paragraph{Sycophantic convergence (M3).}
Agents may adopt each other's \emph{answers} without adopting the underlying reasoning---a form of answer-level sycophancy~\citep{yao2026sycophancy}.
This produces answer convergence (the system reports consensus) while reasoning chains remain divergent or become vaguer.
The high M3 r1 undefined rate---$15.2\%$ on D1 and $14.2\%$ on D2---shows that for a substantial minority of questions, agents' positions are sufficiently incoherent that no majority emerges even after revision.
The zero-step phenomenon (Appendix~\ref{app:undefined}) is the limiting case: an agent produces an answer with no extractable reasoning, indicating that revision has degenerated into pure imitation.

\paragraph{Format effect (\gdp, pre-debate).}
The \textsc{Claim}+\textsc{Ground} structure forces agents to decompose reasoning into named, verifiable steps.
This produces more comparable reasoning units across agents even before any debate interaction: on D1, GDP$_{\text{r0}}$ vs.\ M4$_{\text{r0}}$ alone yields a format-effect $d{=}{+}0.62$ with SIM increase of $+0.034$.
This isolates a structural cause for misalignment in standard debate: a substantial portion of apparent misalignment originates from underspecified output format rather than from genuine reasoning disagreement.

\paragraph{Debate interaction effect (\gdp, r0$\to$r1).}
The \textsc{Stance} mechanism in Round 1 requires each agent to explicitly address every claim made by the other agents.
When an agent must state ``I \textsc{Agree} with Agent B's claim that renal function is preserved, because\ldots'' or ``I \textsc{Disagree} because\ldots,'' the resulting response necessarily shares reasoning vocabulary and logical structure with the other agents.
This drives SIM upward ($+0.077$ on D1, $+0.078$ on D2 from r0 to r1 under \gdp) at a magnitude $2.4{\times}$ larger than the format effect alone ($d{=}{+}1.51$ vs.\ $d{=}{+}0.62$), indicating that \textsc{Stance}-mediated engagement is the primary mechanism.
The format effect is a necessary precondition, however: without structured claims to reference, the \textsc{Stance} mechanism has no well-defined targets.

\section{Cross-Dataset and Cross-Backbone Discussion}
\label{app:crossdataset_extended}

\begin{table}[h]
\centering
\small
\setlength{\tabcolsep}{3pt}
\begin{tabular}{@{}lccc@{}}
\toprule
\textbf{Comparison} & \textbf{Llama D2} & \textbf{Qwen D2} & \textbf{Dir.} \\
& \textbf{($N{=}100$)} & \textbf{($N{=}499$)} & \\
\midrule
GDP$_{\text{r1}}$ vs.\ M3$_{\text{r1}}$ & $+1.99$ & $+1.62$ & \checkmark \\
M3$_{\text{r1}}$ vs.\ M4$_{\text{r0}}$ (illusion) & $-1.32$ & $-0.30$ & \checkmark \\
M3 debate $\Delta$\,HYB (r0$\to$r1) & $-0.038$ & $-0.014$ & \checkmark \\
GDP debate $\Delta$\,HYB (r0$\to$r1) & $+0.014$ & $+0.040$ & \checkmark \\
Format effect $\Delta$\,HYB & $+0.006$ & $+0.017$ & \checkmark \\
\bottomrule
\end{tabular}
\caption{Cross-backbone validation on D2 (referenced from Section~\ref{sec:results:backbone}).
First two rows give Cohen's $d$; remaining rows give $\Delta$\,\carahyb.
All five comparisons are direction-consistent (``Dir.''~\checkmark).}
\label{tab:backbone}
\end{table}

\paragraph{\gdp effect generalization.}
The \gdp effect replicates across all tested conditions with Tier~A effect sizes:
D1 Qwen ($N{=}500$): $d{=}{+}1.43$, $\mathrm{CI}{=}[{+}1.29, {+}1.60]$;
D2 Qwen ($N{=}499$): $d{=}{+}1.62$, $\mathrm{CI}{=}[{+}1.48, {+}1.78]$;
D2 Llama ($N{=}100$): $d{=}{+}1.99$, $\mathrm{CI}{=}[{+}1.65, {+}2.44]$.
The rank ordering of systems is identical across all three conditions:
$\text{M3-GDP}_{r_1}
 \!\succ\! \text{M3-GDP}_{r_0}
 \!\succ\! \text{M4}_{r_0}
 \!\approx\! \text{M3}_{r_0}
 \!\succ\! \text{M3}_{r_1}$
This consistency across two structurally different datasets (fixed 4-option MCQ vs.\ variable 3--10-option clinical reasoning) and two independent backbone models supports the generalizability of both the consistency-illusion finding and the \gdp repair effect.

\paragraph{Consistency-illusion magnitude variation.}
The consistency illusion magnitude varies more across conditions: $d{=}{-}0.08$ (D1, Qwen), $d{=}{-}0.30$ (D2, Qwen), $d{=}{-}1.32$ (D2, Llama).
Two factors plausibly drive this variation.
First, D1's fixed 4-option format constrains the space in which reasoning can diverge while answers remain matched, weakening the measurable illusion compared to D2's mean ${\sim}7.5$-option open answer space.
Second, M3~r1 undefined rates differ markedly across conditions ($15.2\%, 14.2\%, 2.0\%$), and a lower undefined rate preserves more of the worst-alignment cases in the measured sample, strengthening the apparent illusion magnitude.
The third dataset (Llama D2) has the lowest undefined rate and the strongest illusion---an interpretive asymmetry we surface honestly via the sensitivity analysis in Appendix~\ref{app:sensitivity} rather than treating the Llama magnitude as a backbone-effect comparison.

\paragraph{Implications for deployment.}
Where the underlying multi-agent question allows substantial reasoning-space variation (open-ended diagnosis, differential reasoning, treatment selection with multiple acceptable plans), the consistency illusion is likely to be \emph{larger} than on constrained MCQ benchmarks, and the \gdp benefit correspondingly more valuable.
The D1 vs.\ D2 magnitude difference is therefore not a defect of the metric or the intervention but a direct reflection of how much room the task gives reasoning to diverge.

\end{document}